\documentclass{svproc}
\usepackage{url}

\usepackage{graphicx}
\usepackage{mathtools}
\usepackage{xcolor}
\usepackage{cancel}
\usepackage{amssymb}
\usepackage{mathrsfs}
\usepackage{soul}

\newcommand{\snot[1]}{{\scriptstyle \times 10}^{#1}}

 \def\norm#1{\vert #1 \vert}

\begin{document}
\mainmatter              
\title{Arnold diffusion and Nekhoroshev theory}
\titlerunning{Arnold diffusion and Nekhoroshev theory}  
%
\author{Christos Efthymiopoulos \and Roc\'io Isabel Paez}
\authorrunning{Efthymiopoulos \& Paez} 
%
\tocauthor{Christos Efthymiopoulos and Roc\'io Isabel Paez}
\institute{Dipartimento di Matematica Tullio Levi-Civita\\ 
Universit\`a degli Studi di Padova\\ 
Via Trieste 63 35121 Padova, Italy\\ 
\email{cefthym@math.unipd.it, paez@math.unipd.it}\\}

\maketitle              

\begin{abstract}
Starting with Arnold's pioneering work~\cite{Arnold64}, the term
``Arnold diffusion'' has been used to describe the slow
diffusion taking place in the space of the actions in Hamiltonian
nonlinear dynamical systems with three or more degrees of freedom. The
present text is an elaborated transcript of the introductory course
given in the Milano I-CELMECH school on the topic of Arnold diffusion
and its relation to Nekhoroshev theory. The course introduces basic
concepts related to our current understanding of the mechanisms
leading to Arnold diffusion. Emphasis is placed upon the
identification of those invariant objects in phase space which drive
chaotic diffusion, such as the stable and unstable manifolds emanating
from (partially) hyperbolic invariant objects. Besides a qualitative
understanding of the diffusion mechanisms, a precise quantification of
the speed of Arnold diffusion can be achieved by methods based on
canonical perturbation theory, i.e. by the construction of a suitable
normal form at optimal order. As an example of such methods, we
discuss the (quasi-)stationary-phase approximation for the selection
of remainder terms acting as driving terms for the diffusion. Finally,
we discuss the efficiency of such methods through numerical examples
in which the optimal normal form is determined by a computer-algebraic
implementation of a normalization algorithm.  
\end{abstract}
%

\section{Introduction}

In some introductory texts (see, for example,
\cite{rasb-90}\cite{lichlieb-94}\cite{cont-02}), the topic of Arnold
diffusion is introduced by a simplified topological argument, related
to a difference between the cases of invariant tori in Hamiltonian
systems with $n=2$ and with $n\geq 3$ degrees of freedom. Consider a
$n-$degrees of freedom Hamiltonian system $H(q,p)$,
$q\in\mathbb{R}^{n}$, $p\in\mathbb{R}^{n}$, whose phase space contains
a large measure of $n-$dimensional Kolmogorov - Arnold - Moser (KAM)
tori (\cite{Kolmo-54,Arnold-63,Moser-62}).  Any orbit $(q(t),p(t))$
with initial conditions $(q_0,p_0)$ on a KAM torus remains forever
confined to the torus. Any orbit with initial conditions $(q_0,p_0)$
belonging to the {\it complement}, in phase space, with respect to the
set of all KAM tori, remains confined to the $(2n-1)$-dimensional
manifold defined by the orbit's constant energy value ${\cal
  M}_E:=\{(q,p)\in\mathbb{R}^{2n}: H(q,p)=E = H(q_0,p_0)\}$.  Take
first $n=2$. Thus, $\dim\left({\cal M}_E\right) = 3$. Suppose there is
a KAM torus ${\cal T}$ embedded in the same energy manifold. We have
$\dim\left({\cal T}\right)=2$.  Since the torus's dimension differs
just by one from the dimension of the energy manifold, ${\cal T}$
divides ${\cal M}_E$ into two parts, which can be called the
`interior' and the `exterior' of the torus. Furthermore, since
$H(q,p)$ is autonomous, there can be no trajectory going from the
interior to the exterior of the torus; such a trajectory would
necessarily have to cross transversally the torus at a point
$(q(t_c),p(t_c)) \in{\cal T}$ at some time $t_c$, but this is
impossible since the flow on the torus is invariant, i.e., the initial
condition $q=q(t_c),p=p(t_c)$ would lead necessarily to a trajectory
confined on the torus. We roughly refer to this as the `dividing
property' of KAM tori in systems with $n=2$ degrees of freedom. On the
other hand, there is no dividing property of the KAM tori when $n\geq
3$, since, in that case $\dim\left({\cal M}_E\right) -\dim\left({\cal
  T}\right)\geq 2$. For example, when $n=3$ we have $\dim({\cal
  M})=5$, and $\dim({\cal T})=3$, thus ${\cal T}$ cannot divide ${\cal
  M}$ into disconnected sets. To visualize this just lower all
dimensions in the above examples by one: hence, a circle (dimension 1)
divides a plane (dimension 2) to the interior and the exterior of the
circle, while a circle embedded in Euclidean space (dimension 3)
cannot divide the latter into disconnected sets.

The non-existence of topological barriers when $n\geq 3$ renders {\it
  a priori} possible to have long excursions of the chaotic orbits
throughout the whole constant energy manifold.  However, two questions
become immediately relevant: i) can we prove that the chaotic orbits
{\it do really undergo} those (topologically allowed) arbitrarily long
chaotic excursions?  ii) is the timescale involved short enough to
make the phenomenon relevant and worth of further study as regards
applications in physical systems (including, for the purposes of the
present course, systems of interest in celestial mechanics or
astrodynamics)?

We refer to question (i) above as the problem of the {\it existence of 
Arnold diffusion}. In the words of Lochak's influential review~\cite{Lochak-99}, 
it is the problem of demonstrating that ``topological transitivity on the 
energy surface generically takes place''.  We refer, instead, to question 
(ii) as the problem of how to quantitatively estimate the {\it speed of 
Arnold diffusion}. Addressing this question in the context of particular 
problems encountered in physics and astronomy requires a (partly heuristic) 
use of computational techniques, as discussed in detail, for example, 
in a well known review by Chirikov~\cite{chiri-79}. It is worth mentioning that, 
after about 60 years of research, only partial answers are available today 
regarding both questions. In particular, the existence of Arnold diffusion has 
been rigorously established in various cases of so-called {\it a priori unstable} 
systems (see \cite{chiergal-94}\cite{cheyan-04}\cite{gidllave-06}\cite{delshetal-16}). 
Instead, it remains an open problem in the far more difficult case of {\it a priori 
stable} systems (see section 3 for definitions). In the latter case, we avail, 
however, ample numerical evidence of the global drift of the trajectories within the 
so-called Arnold's {\it web of resonances}, as visualized in a series of beautiful 
numerical studies (\cite{legaetal-03}\cite{guzzoetal-05}\cite{Frosetal-05}
\cite{guzzoetal-06}; see \cite{legaetal-07} for a review). 
In fact, the visualization of the Arnold web in a priori stable systems was made 
possible by the use of techniques allowing to carefully choose initial conditions 
along the thin resonance layers in phase space marked by the web of resonances. 
The Fast Lyapunov Indicator (FLI,~\cite{Frosetal-97}) is an example of such 
technique (see also the lecture by M. Guzzo in the present volume of proceedings).

As emphasized by Lochak~\cite{Lochak-99}, a demonstration that the Arnold
diffusion really takes place requires establishing the existence of a
{\it mechanism of transport} for the weakly chaotic orbits within the
Arnold web. Arnold's original example~\cite{Arnold64} actually describes
such a mechanism. This is based on proving the existence of
heteroclinic intersections between the stable and unstable manifolds
emanating from a set of nearby partially hyperbolic low-dimensional tori
arranged in a so-called `transition chain'. One initially demonstrates
that two nearby tori, of a small distance, say, ${\cal O}(\delta)$,
where $\delta$ is a small parameter, exhibit a `splitting of the
separatrices' (their stable and unstable manifolds) such that these
manifolds develop heteroclinic intersections. Let $\tau_i$,
$i=1,2,\ldots$ be a sequence of tori, $\tau_i$ being neighbor to
$\tau_{i-1},\tau_{i+1}$. Assume we know that the unstable manifold
emanating from $\tau_i$ has a heteroclinic intersection with the
stable manifold ending at $\tau_{i+1}$. Then, there is a `doubly
asymptotic' orbit which tends to $\tau_i$ as $t\rightarrow-\infty$,
while it tends to $\tau_{i+1}$ forward in time as $t\rightarrow\infty$. 
Such orbits can be established for any pair $\tau_i,\tau_{i+1}$, 
$i=1,2,...$, but of course they cannot themselves be the orbits of 
Arnold diffusion, since they never go very far either from $\tau_i$ 
or $\tau_{i+1}$.  On the other hand, invoking a so-called `shadowing 
lemma' (see \cite{delshetal-08} for a review), one demonstrates that there 
are true orbits of the system which shadow the whole chain of heteroclinic
orbits established in the above way. Thus, these shadowing orbits
undergo Arnold diffusion. A quick estimate of the speed of diffusion
is obtained as follows: upon completion of one cycle of the transition
mechanism, the trajectory has traveled a distance $S={\cal O}(\delta)$
in a time $T_{i,i+1}\approx T_s$, which roughly coincides with the
time required to cover one homoclinic loop close to the separatrix of
the resonance associated with the unstable tori $\tau_i$ (see section
2 below). Hence, the local speed of Arnold diffusion is
$V_{AD}\approx\delta/T_s$, where both parameters $\delta$ and $T_s$
depend on the small parameters of the problem under study. Of course,
this is an oversimplified estimate. Estimates of practical interest
are rather hard to obtain, as explained in the sections to follow. On
the other hand, the topic of how to describe itself the one-step
transition of the chaotic trajectories far from, and then back to the
asymptotic ends of the intersecting manifolds has been developed
substantially in recent years, leading to the concept of the so-called
`scattering map' (see \cite{delshetal-06}\cite{delshetal-08b}). 
Applications of the scattering map technique in Celestial Mechanics 
are discussed, in particular, by~\cite{canaletal-06} (see also 
\cite{Capetal-17} and references therein).

Regarding numerical investigations of Arnold diffusion, since this is
a slow phenomenon its revelation requires a rather high computing
power and the capacity to numerical propagate large sets of
trajectories over long integration times. Owing to its complexity, the
numerical investigation of the weakly chaotic diffusion has so
far been limited to few DOF dynamical systems, including several 
systems of particular interest for dynamical astronomy (see an 
extensive, but only indicative, list of references in section 4 
of \cite{LaPlata}). However, it is unclear whether the notion of 
Arnold diffusion can be useful for the analysis of the diffusive 
processes in all those models. On the other hand, there are cases,  
in particular around normally hyperbolic invariant objects in the 
restricted three-body problem, where Arnold diffusion has been explicitly 
demonstrated to apply (see, for example, \cite{moeckel-96}\cite{canaletal-06}
\cite{Capzgli-11}\cite{Capetal-17}\cite{fejozetal-16}).

The present tutorial is organized as follows: section 2 presents in
some detail the original example discussed in~\cite{Arnold64}, serving
to introduce most elements of the conceptual framework for the
discussion of Arnold diffusion. Section 3 deals with the case of a
priori stable systems and with the connection of Arnold diffusion with
Nekhoroshev theory. Finally, Section 4 discusses various
semi-analytical approaches to the quantification of the speed of
Arnold diffusion.

\section{Arnold's example}
The Hamiltonian model presented by Arnold in~\cite{Arnold64} is 
\begin{equation}\label{eq:arnold2half}
  H(q,\phi_1,t,p,J_1) = \frac{1}{2} p^2 + \frac{1}{2} J_1^2 + \epsilon
  \left(\cos q - 1 \right)\, \left(1 + \mu \left( \sin \phi_1 + \cos t
  \right) \right)~,
\end{equation}
It is a model of a pendulum (variables $(q,p)$) coupled with a rotator (variables 
$(\phi_1,J_1)$) via the time-dependent term $\epsilon\mu\cos q\cos t$. 
We will assume $\epsilon>0$ and fixed, while varying $\mu$, with $|\mu|<<\epsilon$. 
The Hamiltonian can be formally extended to 3DOF autonomous by introducing the angle 
$\phi_2 = t$ conjugated to a dummy action $J_2$:
\begin{equation}\label{eq:arnold3d}
H\rightarrow  H(q,\phi_1,\phi_2,p,J_1,J_2) = \frac{1}{2} p^2 + \frac{1}{2} J_1^2 +
  J_2 + \epsilon \left(\cos q - 1 \right)\, \left(1 + \mu \left( \sin
  \phi_1 + \cos \phi_2 \right) \right)~.
\end{equation}

For $\mu = 0$, we have $\dot{J_1} = \dot{J_2} = 0$, thus the actions remain invariant 
along the trajectories. For any values $(J_1,J_2)$, the angles $\phi_1$, $\phi_2$ evolve 
linearly with frequencies $\omega_1 = J_1$, $\omega_2 = 1$. Thus, changing the value of 
$J_1$, we can obtain any desired frequency ratio $\omega_1/\omega_2 = J_1$ (the dummy 
action $J_2$ can be set initially to any value (e.g. $J_2(0)=0$) without consequences 
for the dynamics). 

Consider now the case $\mu \neq 0$. For generic trajectories, we obtain $\dot{J}_1\neq 0$, 
$\dot{J}_2\neq 0$. However, there is a particular set of initial conditions for which 
the trajectories preserve the actions: 
\begin{equation}\label{eq:hyptorus}
\tau(J_1,J_2)=\left\{
q=p=0, J_1=const,J_2=const,(\phi_1,\phi_2)\in\mathbb{T}^2
\right\}~~.
\end{equation}
Taking Hamilton's equations for the complete system:
\begin{eqnarray}\label{eq:eqmoarn}
\dot{q}&=p,~~~~\dot{p} &= -\epsilon\sin q(1+\mu(\sin\phi_1+\cos\phi_2)) \nonumber\\
\dot{\phi_1}&=J_1,~~~\dot{J}_1 &=~~\epsilon\mu(\cos q-1)\cos\phi_1 \\
\dot{\phi_2}&=1,~~~\dot{J}_2 &= -\epsilon\mu(\cos q-1)\sin\phi_2 \nonumber
\end{eqnarray}
we immediately find that any initial condition in the set
$\tau(J_1,J_2)$ leads to $\dot{q}=\dot{p}=0=\dot{J}_1=\dot{J}_2 = 0$,
while $\dot{\phi}_1=J_1(t)=const$, $\dot{\phi}_2=J_2(t)=const$. Thus,
$\tau(J_1,J_2)$ is invariant under the flow and homeomorphic to the
2D-torus $(\phi_1,\phi_2)\in\mathbb{T}^2$. We will denote by
$\mathcal{T}$ the invariant set formed by the family of all the tori
$\tau(J_1,J_2)$ $(J_1,J_2)\in\mathbb{R}^2$.

The invariance of the tori $\tau(J_1,J_2)$ crucially relies on having set  
$(q,p)$ as $(q,p) = (0,0)$. We now wish to explore what will happen if, 
instead, we choose the initial condition $(q_0,p_0)$ close to, but not equal to 
$(0,0)$. For example, we can set $q_0=0, p_0\neq 0$, with $|p_0|<D$ and $D$ 
small, and $(J_{1,0},J_{2,0},\phi_{1,0},\phi_{2,0})$ chosen at will. We then 
want to understand the future evolution, in particular of the actions $J_1(t),
J_2(t)$, as a consequence of choosing initial conditions in the neighborhood 
of, but not exactly on the torus $\tau(J_{10},J_{20})$. Addressing this question 
requires the use of a mixture of analytical as well as geometric arguments. 
Let us summarize some basic ones:

\subsection{Existence of KAM tori}
We can demonstrate the existence of Kolmogorov-Arnold-Moser (KAM) tori for a 
Cantor set (of non-zero measure) of initial conditions $p_0$ along the line 
$q=0$. Decomposing the Hamiltonian as:
\begin{equation}\label{eq:hamarnkam}
H(q,\phi_1,\phi_2,p,J_1,J_2) = H_0(p,J_1,J_2)+\epsilon H_1(q,\phi_1,\phi_2,p,J_1,J_2;\mu) 
\end{equation}
where $H_0 = {1\over 2}(p^2+J_1^2) + J_2$, $H_1 = \epsilon \left(\cos
q - 1 \right)\, \left(1 + \mu \left( \sin\phi_1 + \cos \phi_2 \right)
\right)$, the Hamiltonian $H_0$ satisfies the iso-energetic
non-degeneracy condition:
\begin{equation}\label{eq:isodeg}
\det\left(
\begin{array}{cc}
\mbox{Hess}(H_0) & \nabla_I(H_0)\\
(\nabla_I(H_0))^T & 0
\end{array}
\right)=0
\end{equation}
where $\mbox{Hess}(H_0)$ is the $3\times 3$ Hessian matrix of $H_0$ with respect to 
$I\equiv(p,J_1,J_2)$. Thus, the necessary conditions for the Kolmogorov theorem~\cite{Kolmo-54}
hold, namely:\\
\\
\noindent
{\bf Theorem (Kolmogorov 1954):} there exist positive constants $\epsilon_0,\gamma,\tau$ 
such that, for $|\epsilon|<\epsilon_0$, and $(p_0,J_{10})$ such that the frequencies 
$\omega_p=(\partial H_0/\partial p)_{p=p_0} = p_0$, $\omega_1=J_1$, $\omega_2=1$ satisfy 
the Diophantine condition
\begin{equation}\label{eq:dioph}
|k_p\omega_p+k_1\omega_1+k_2\omega_2|>{\gamma\over k^\tau} 
\end{equation}
where $k=|k_p|+|k_1|+|k_2|$, the trajectory with initial conditions
$p(0)=p_0$, $J_1(0)=J_{10})$, $q(0)=0$, $J_2(0)=J_{20}\in\mathbb{R}$,
as well as $(\phi_1(0),\phi_2(0))\in\mathbb{T}^2$ lies in a
three-dimensional torus, where all phase-space co-ordinates evolve
quasi-periodically with the frequencies
$(\omega_p,\omega_1,\omega_2)$. \\ \\
The above theorem can be proven
by the construction of the so-called \textit{Kolmogorov normal form}
in the neighborhood of the chosen initial conditions.  The value of
$\gamma$ restricts the measure of initial conditions satisfying the
Diophantine condition. By number-theoretical arguments we find
$|p_0|>D={\cal O}(\gamma)$, hence motions very close to $p_0=0$ cannot
be quasi-periodic.

\subsection{Semi-analytical (`Melnikov') approach}
\begin{figure}[!ht]
\centering
\includegraphics[width=1.\textwidth]{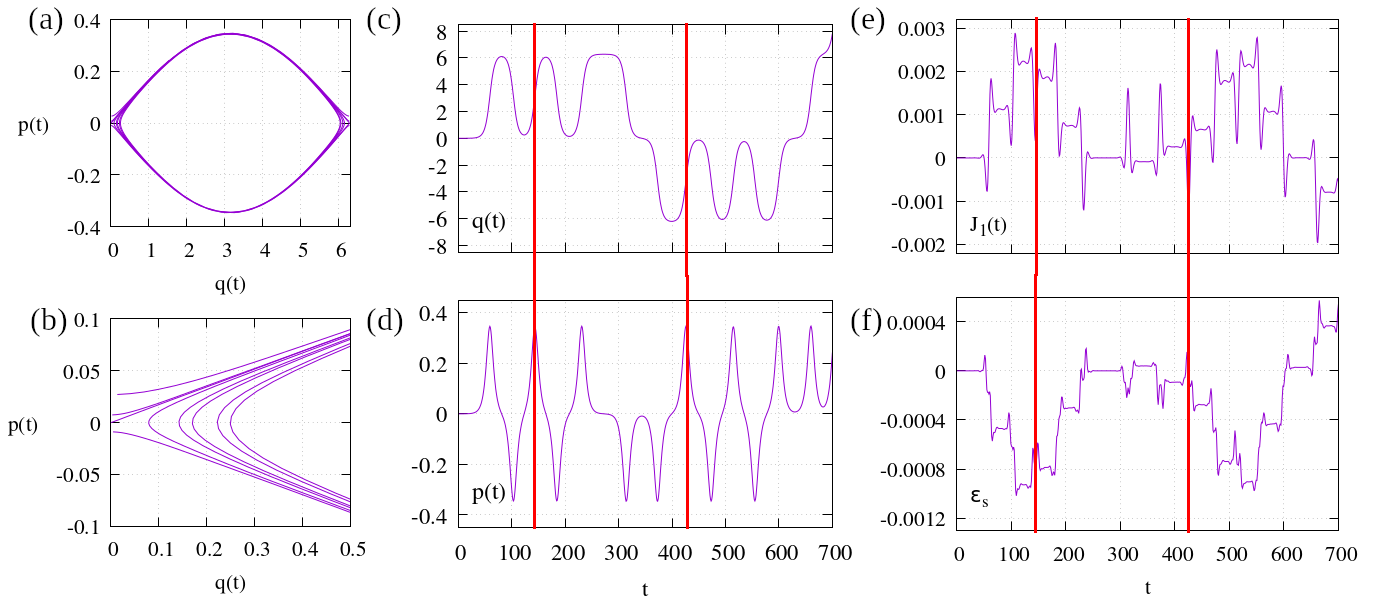}
\caption{Evolution of the orbit with initial conditions $\phi_1(0)=0$,
  $\phi_2(0)=0$, $p(0) = 5 \snot[-5]$, $J_1(0) = 0.3\sqrt{2}$, $J_2 =
  1$, under the flow of the Hamiltonian~\eqref{eq:arnold3d} with
  $\epsilon = 0.03$ and $\mu =0.01$: $p(t)$ vs $q(t)$ in panel (a) and
  (b), $q(t)$ in panel (c),$p(t)$ in panel (d), $J_1(t)$ in panel (e),
  in the time interval $t\in [0,700]$. (f) Evolution of the pendulum
  energy $\varepsilon_s$ for the same orbit. In (a) the angle 
  $q(t)$ is shown modulo $2\pi$. The two red vertical lines in panels (c) 
  to (f) are helping guides to the eye: they indicate two different moments 
  where the trajectory passes from the uppermost point of the separatrix. 
  All jumps in $J_1(t)$ occur at these passages.}
\label{fig:arnoldex}
\end{figure}

In order to deal with non-quasiperiodic motions, very close to the
torus $p_0=0$, we can try to approximate the evolution of the variables
$(\phi_1,\phi_2,J_1,J_2)$ by a model in which the evolution in the
variables $(q(t),p(t))$ is \textit{a priori} modeled via some
`near-separatrix' analytical approximation
$(q_s(t;\varepsilon_s),p_s(t;\varepsilon_s))$ based on the pendulum
model (or, in general, the model of resonance giving rise to a particular form 
of the separatrix). This strategy is explored heuristically in a well known 
review on Arnold diffusion by Chirikov~\cite{chiri-79} and set in a rigorous base 
in \cite{holmmars-82}. It is based on the remark that choosing $(q_0,p_0)$ 
very close to the values $(0,0)$ leads to a motion in the variables $(q(t),p(t))$ 
which can be modeled as a sequence of stochastic alterations between pendulum 
librations or rotations, each with nearly conserved pendulum energy
\begin{equation}\label{eq:pendene}
h_s(q,p)=\varepsilon_s = {1\over 2}p^2 + \epsilon(\cos q -1)~~
\end{equation}
with $\varepsilon_s\approx\varepsilon_{s,0}=0$ (corresponding to the
invariant torus $(q,p)=(0,0)$). Figure~\ref{fig:arnoldex} exemplifies
this approach. The figure shows the evolution of the trajectory with
initial conditions $q(0)=0$,$\phi_1(0)=0$, $\phi_2(0)=0$, $p(0) = 5
\snot[-5]$, $J_1(0) = 0.3\sqrt{2}$, $J_2 = 0$, under the {\it
  complete} flow (\ref{eq:arnold3d}), with $\epsilon = 0.03$ and $\mu
= 0.01$.
Since the coupling term between pendulum and the rest of the system
has size ${\cal O}(\mu\epsilon)$, with $\mu=0.01$ this term is two
orders of magnitude smaller than the $\epsilon\cos q$ term defining
the pendulum separatrix. As a consequence, the `splitting' of the
separatrix will be quite small. This means that there will be only a
small error in approximating the evolution of $(q(t),p(t))$ \textit{as
  if it was governed only by the pendulum Hamiltonian} $h_s(q,p)$
(Eq.(\ref{eq:pendene})). Figure \ref{fig:arnoldex} indicates that this
is essentially correct. Denote by $R(+),R(-)$ a pendulum rotation with
the Hamiltonian $h_s(q,p)$ and with $p>0$ or $p<0$ respectively, and
by $L(+),L(-)$ the upper and lower parts (again $p>0$ or $p<0$) of a
librational curve in the same Hamiltonian. Then, the evolution of
$p(t),q(t)$ in Fig.~\ref{fig:arnoldex} can be represented as a
sequence of segments of pendulum librational or rotational curves.  Up
to $t=700$ we have
$$
R(+),L(-),L(+),L(-),L(+),R(-),R(-),L(+),L(-),L(+),L(-),R(+),R(+),\ldots
$$
Denoting by $T_{s,i}$, $i=1,2,\ldots$ the time it takes to
accomplish one segment, the times $T_{s,i}$ can be estimated as the
times between two successive local extrema in
Fig.~\ref{fig:arnoldex}(c). We find that $T_{s,i}$ has value nearly
always around $T_s\lesssim 100$. Also, using the values
$q(t_i),p(t_i)$ at the times $t_i$ of the local extrema of the curve
$q(t)$, we can compute a sequence of corresponding pendulum energies
$\varepsilon_i = h_s(q(t_i),p(t_i))$ characteristic of each segment.

Chirikov~\cite{chiri-79} proposed a model to study the qualitative
properties of the mapping
$(q(t_i),p(t_i))\rightarrow(q(t_{i+1}),p(t_{i+1}))$, or, equivalently,
$\varepsilon_i\rightarrow\varepsilon_{i+1}$, $t_i\rightarrow t_{i+1}$,
called, by him the {\it whisker mapping} (`whiskers' meaning the
separatrices of the torus $(q,p)=(0,0)$). Figure~\ref{fig:arnoldex}(f)
shows the first few transitions in the energy values $\varepsilon_s$. 
In every step, $\varepsilon_s(t)$ takes nearly constant value in a `plateau', 
separated from the next plateau by a rapid oscillation. These oscillations 
take place mid-way along each homoclinic transition far from and back to 
the neighborhood ofthe torus $(q,p)=(0,0)$.

We now discuss how to exploit the above empirical information in order to model the 
evolution in the remaining variables $J_1$, $J_2$, $\phi_1$, $\phi_2$ along such 
homoclinic transitions. The so-called `Melnikov approach' consists essentially of the 
following approximation: in the interval $t_i<t<t_{i+1}$, we will evolve 
the remaining variables according to the approximate system 
\begin{eqnarray}\label{eq:eqmomel}
\dot{\phi_1}&=J_1,~~~\dot{J}_1 &=~~\epsilon\mu(\cos q_s(t)-1)\cos\phi_1 \\
\dot{\phi_2}&=1,~~~\dot{J}_2 &= -\epsilon\mu(\cos q_s(t)-1)\sin\phi_2 \nonumber
\end{eqnarray}
which is the same as the original system but with $q(t),p(t)$ substituted with by the 
solutions $q_s(t),p_s(t)$ of the pendulum equations
\begin{equation}\label{eq:pendeq}
\dot{q}_s=p_s,~~~~\dot{p}_s = -\epsilon\sin q_s
\end{equation}
with initial conditions $q_s=q(t_i),p_s=p(t_i)$.
\begin{figure}[!ht]
  \centering
  \includegraphics[width=1.\textwidth]{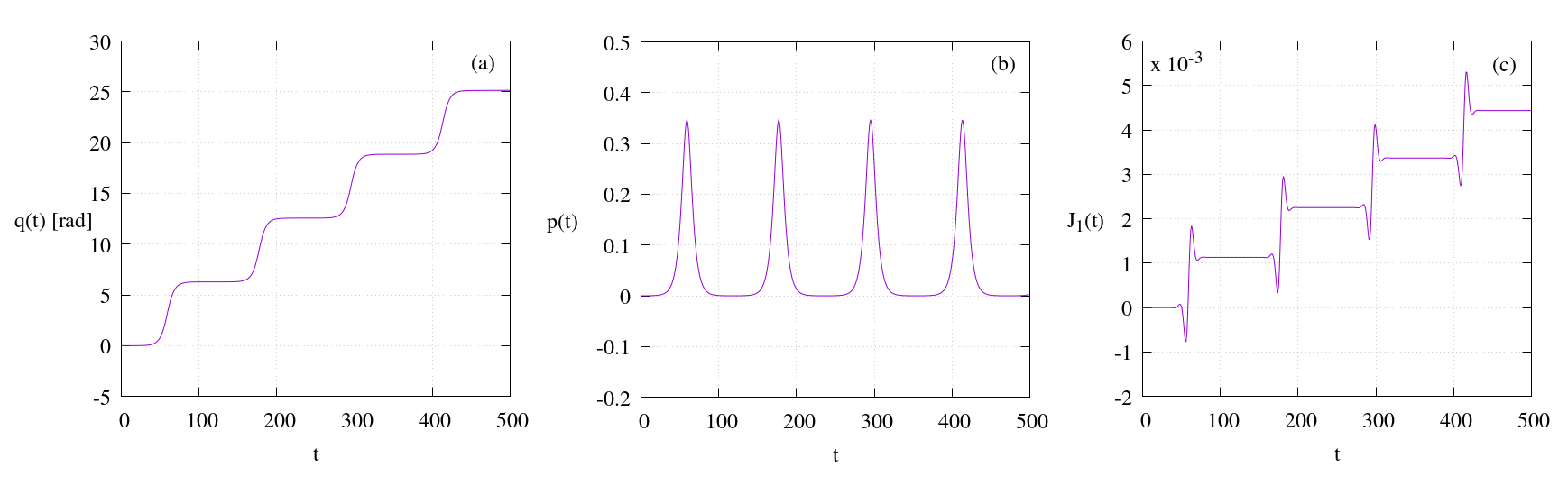}
  \caption{Evolution of the pendulum solution (a) $q_s(t)$, (b) $p_s(t)$, for the 
  same initial condition as in Fig.\ref{fig:arnoldex}, namely $q_s(0)=0$, $p_s(0)=
  5\times 10^{-5}$, but following the pendulum equations (Eqs.(\ref{eq:pendeq})). 
  (c) Evolution of the action $J_1(t)$ under the equations of the Melnikov approximation 
  (Eqs.(\ref{eq:eqmomel})). We observe that $J_1(t)$ exhibits jumps in time which are 
  qualitatively similar as those of the complete model, shown in panel (e) of 
  Fig.\ref{fig:arnoldex}.}
  \label{fig:melnikov}
\end{figure}

Figure~\ref{fig:melnikov} shows the evolution under the approximate
equations (\ref{eq:pendeq}) and (\ref{eq:eqmomel}), starting with the
same initial condition as in Fig.~\ref{fig:arnoldex}, which belongs to
the upper rotation domain of the pendulum $(q(0)=0,p(0)>0)$. Since we
now integrate the exact pendulum equations we obtain a periodic
evolution of the angle $q$ completing a circle at the period
$T(\varepsilon_s)$ given by
\begin{equation}\label{eq:pendper}
T_s(\varepsilon_s) \simeq
\int_{0}^{2\pi} {dq\over\sqrt{2\left(\varepsilon_s-\epsilon(\cos q -1)\right)}}
= {32\over\sqrt{\epsilon}}\ln\left({|\varepsilon_s|\over\epsilon}\right)~~.
\end{equation}
However, the action variable $J_1(t)$ (Fig.\ref{fig:melnikov}(c))
undergoes abrupt jumps of size $~10^{-3}$ every time when the pendulum
variables are mid-way along accomplishing one homoclinic transition.

\begin{figure}[!ht]
  \centering
  \includegraphics[width=1.0\textwidth]{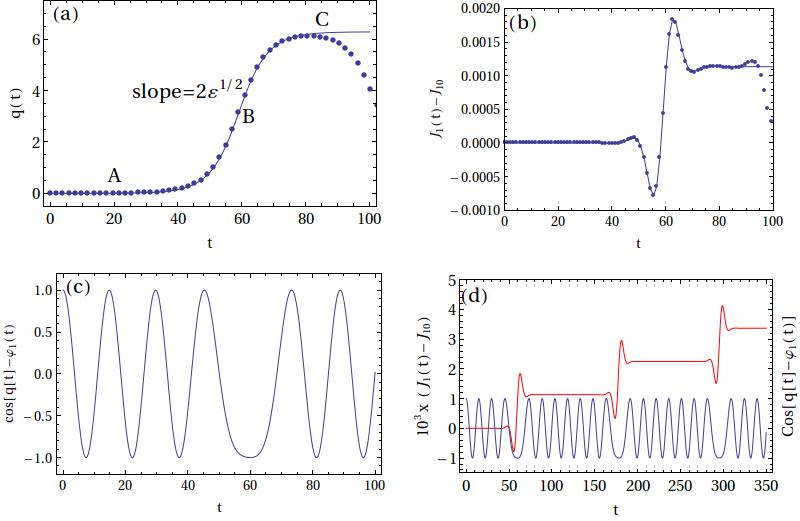}
  \caption{(a) The evolution of the variable $q(t)$ along the first
    homoclinic transition, as obtained by numerical integration in the
    complete model (\ref{eq:eqmoarn}) (points), and with the model of
    Eq.(\ref{eq:qstsep}) (solid curve). (b) The first observed
    numerical jump in $J_1(t)$ (points) against the prediction of the
    model of Eq.(\ref{eq:j1tmel}). (c) The curve
    $\cos(q(t)-\phi_1(t))$ in the time interval corresponding to the
    first jump. (d) Left axis: several jumps in the variable $J_1(t)$
    compared with (right axis) the evolution of
    $\cos(q(t)-\phi_1(t))$. The jumps take place at precisely those
    points where the phase $q-\phi_1$ forms a local plateau, departing
    from a pure oscillation.}
  \label{fig:melnjump}
\end{figure}

The jumps in Fig.~\ref{fig:melnikov}(c) are qualitatively quite
similar to the jumps seen in the real orbit
(Fig.~\ref{fig:arnoldex}(e)). In fact, the real jumps can be easily
modeled by one further simplification: since all along the depicted
solution $J_1(t)$ undergoes only a small (${\cal O}(10^{-3})$)
variation around the initial value $J_{10}=0.3\sqrt{2}$, we can
approximate the solution of the angular equation
$\dot{\phi_1}(t)=J_1(t)$ by $\phi_1(t) = \phi_{1,0}+J_{10} (t-t_0)$,
where $\phi_{1,0}$ is the value of the angle $\phi_1$ at the starting
time $t_0$ of one homoclinic transition.  We also approximate the
solution $q_s(t)$ by the one holding along the pendulum separatrix:
\begin{equation}\label{eq:qstsep}
q_s(t) \approx 4\arctan\left(e^{\sqrt{\epsilon}(t-t_0-T_s/2)}\right)~~,
\end{equation}
with $T_s$ still given by Eq.~(\ref{eq:pendper}) (this last
approximation is not really needed, but makes the computation easier
with respect to the pendulum solution for the exact initial conditions
given in terms of elliptic functions). As shown in
Fig.~\ref{fig:melnjump}(a), the separatrix solution (\ref{eq:qstsep})
fits the evolution of $q(t)$ along the first homoclinic transition as
obtained numerically by the complete model (\ref{eq:eqmoarn}) up to a
time $t\approx 80$, where the real orbit starts its second homoclinic
transition. Using the above approximations, all quantities in the
r.h.s. of the differential equation for $J_1$ in the system
(\ref{eq:eqmomel}) becomes explicit functions of the time $t$, Then,
the approximative solution $J_1(t)$ can be obtained by quadratures:
\begin{equation}\label{eq:j1tmel}
J_1^{(M)}(t) = J_1(0) + \epsilon\mu\int_0^t  
\left(\cos(4\arctan(\exp(\sqrt{\epsilon}(t'-T_s/2))))-1\right)
\cos(\phi_{10}+J_{10} t')dt'
\end{equation}
An integral of the form (\ref{eq:j1tmel}) is called a `Melnikov
integral'. It has the distinguishing feature that the integrand
contains trigonometric functions $\cos\phi$, with
$\phi=m_qq+m_1\phi_1+m_1\phi_2$, $(m_q,m_1,m_2)\in\mathbb{Z}^3$, for
some of which the evolution is not linear in time, as for example, the
angle $q$ which follows the near-separatrix pendulum solution
(\ref{eq:j1tmel}).  Figure~\ref{fig:melnjump}(b) shows the comparison
between the `Melnikov' model $J_1^{(M)}(t)$ and the real evolution of
the same variable up to the end of the first homoclinic transition,
showing an excellent fit for the observed jump of the action $J_1(t)$.

How can we understand this success of the `Melnikov approximation'? Of
course the answer is hidden in the properties of the quadrature
(\ref{eq:j1tmel}). As a coarse remark, by the equation for
$\dot{J}_1(t)$ in (\ref{eq:eqmomel})), the evolution of $J_1(t)$ is
determined by the terms $\cos(q+\phi_1)$, $\cos(q-\phi_1)$ and
$\cos\phi_1$. We saw that $\phi_1$ evolves nearly linearly
$\phi_1(t)\approx\phi_1(0)+J_{10}t$, so the integral
$\int_0^t\cos(\phi_1(t'))dt' \approx {1\over
  J_{10}}\sin(\phi_1(0)+J_{10}t)$ will only produce some rapid
oscillation in the evolution of $J_1(t)$. The remaining terms,
however, $\cos(q+\phi_1)$, $\cos(q-\phi_1)$ depend on the angle $q$,
which evolves approximately by the pendulum trajectory of
Eq.~(\ref{eq:qstsep}) (as shown in Fig.~\ref{fig:melnjump}(a)). Now,
the pendulum trajectory spends most of the time near the unstable
origin, hence we have $\dot{q}\approx 0$ there. On the other hand the
speed $\dot{q}$ in the middle of the homoclinic transition can be
estimated as $\dot{q}(t)\approx 2\sqrt{\epsilon}$ (equal to
$\dot{q}_s(t=T_s/2)$ in Eq.~(\ref{eq:qstsep})). Thus, the curve $q(t)$
consists, essentially, of three parts, marked in
Fig.~\ref{fig:melnjump}(a) by A,B,and C respectively. In the domains
A,C the curve is nearly horizontal, and
$\cos(q\pm\phi_1)\simeq\cos(\phi_1)$, thus the integrals
$\int\cos(q\pm\phi)$ yield essentially the same oscillatory behavior
as for the integral $\cos\phi_1$ alone. In the domain $B$, instead, we
have a slower evolution of the angle $q-\phi_1$: in our example we
have $\dot{q}-\dot{\phi}_1 \simeq 2\sqrt{\epsilon}-J_{10} =
-0.07785...$ in B, compared to $\dot{q}-\dot{\phi}_1
\simeq\sqrt{2\epsilon} = 0.34...$, $J_{10}=0.4242...$ in A or C.  As a
consequence, The curve $\cos(q(t)-\phi_1(t))$ develops an approximate
`plateau' near the time $t=T_s/2\simeq 59.$
(Fig.~\ref{fig:melnjump}(c)). Since the integrand of the Melnikov
integral in (\ref{eq:j1tmel}) temporarily stabilizes to a constant
value, the integral will give a locally linear evolution of $J_1(t)$,
thus causing a quick jump, lasting roughly as the time duration of
B. After exit from B, the $J_1(t)$ returns to an oscillatory
evolution, which keeps up to the next homoclinic transition. More
jumps then occur at each successive homoclinic transition, as shown in
(Fig.~\ref{fig:melnjump}(d)).

\begin{figure}[!ht]
  \centering
  \includegraphics[width=1.0\textwidth]{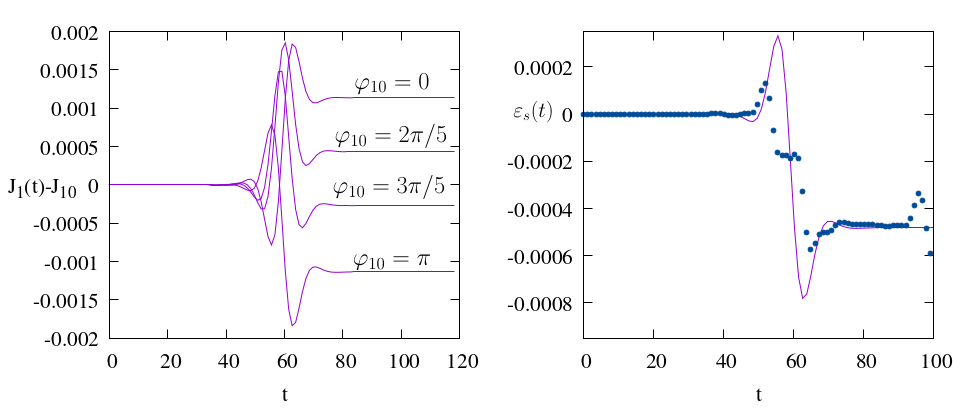}
  \caption{\textit{Left:} the jumps in the variable $J_1$ obtained
    through formula (\ref{eq:j1tmel}) by changing the initial angle
    $\phi_0$ according to the values indicated in the
    figure. \textit{Right:} The jump in the pendulum energy
    $\varepsilon_s=p^2/2+\epsilon(\cos q-1)$ as computed for the
    numerical orbit in the complete model (points) and with the
    `Melnikov model' of Eq.(\ref{eq:enesmel}).}
  \label{fig:ranphase}
\end{figure}
Comparing the above picture with Fig.\ref{fig:arnoldex}(e), we do now
interpret qualitatively the nature of the jumps, but we still need to
understand why the jumps differ in size and/or sign. The sequences of
times where jumps occur can be estimated by $t_{i+1}-t_i\approx
T_s(\varepsilon_{s,i})$, with $T_s$ given by
Eq.~(\ref{eq:pendper}). These times are of similar order, but different
one from the other even for a small change in $\varepsilon_s$ (compare
the times $T_s$ when $\varepsilon_s=10^{-5}$ or $10^{-3}$). As a
consequence, at the starting point of each homoclinic transition, the
orbit is at a different value of the starting angle
$\phi_{1,0}$. However, as shown in Fig.~\ref{fig:ranphase}, according
to the value of $\phi_{1,0}(t_i)$ we can obtain jumps in $J_1$ of
various sizes, positive or negative.  Under the assumption that the
sequence $\phi_{1,0}(t_i)$ is random (`random phase approximation'),
this leads to a random walk model for the variations of $J_1(t)$. In
reality, long correlations can survive in the sequence
$\phi_{1,0}(t_i)$, and the diffusion in $J_1(t)$ can partly loose its
normal character (typically the dynamics becomes sub-diffusive, see
\cite{mestreetal-12}). Also, using the Melnikov approach, we may compute a
continuous in time approximation for the evolution of the energy
$\varepsilon_s(t)$
\begin{equation}\label{eq:enesmel}
\varepsilon_s = \varepsilon-\frac{1}{2} (J_1^{(M)})(t))^2 -J_2^{(M)}(t) 
- \epsilon\mu \left(\cos(q_s(t)) - 1 \right)
\left( \sin(\phi_{10}+J_{10}t) + \cos (\phi_{20}+t) \right)~,
\end{equation}
where $J_2^{(M)}(t)$ is the `Melnikov' model for the evolution of the
action $J_2$, analogous to the model (\ref{eq:j1tmel}) for the action
$J_1$. The right panel in Fig.~\ref{fig:ranphase} shows the evolution
of the pendulum energy $\varepsilon_s(t)$ for the first jump in the
real orbit and as obtained by the model (\ref{eq:enesmel}), showing
again a good fit. Then using all the above approximations, we can
arrive at a heuristic model for Chirikov's `whisker map'. While
deterministic, in practice this model leads to nearly random sequences
$\varepsilon_i$, $\phi_{1,i}$, that is, to a stochastic process for
the evolution of the orbit in the action space. Estimating the value
of the diffusion coefficient relies on some semi-analytical
approaches, as discussed in sections 3 and 4 below.

As a final comment, one can remark that the `plateaus' of the curve
$\cos(q-\phi_1)$, responsible for the jumps Fig.~\ref{fig:melnjump}(d),
are due to the tuning of the values of $\dot{q}$ and
$\dot{\phi}_1\simeq J_{10}$ at region B of Fig.~\ref{fig:melnjump}.
This tuning is rather exceptional, and was essentially imposed for
illustration purposes by the choice of the initial condition
$J_{10}$. Generic initial conditions instead (as, for example,
choosing $J_{10}$ one order of magnitude larger) will destroy such
tuning.  Does this imply that there is no more drift in action space
by jumps as the above?  As will be discussed in section 4, we can make
a number of steps of perturbation theory, seeking to eliminate
altogether the now useless combinations $\cos(q-\phi_1)$,
$\cos(q+\phi)$ and prove perpetual stability for the actions $J_1$ and
$J_2$. However, doing so generates new `dangerous' harmonics along the
normalization process (see, for example, \cite{morbgio-97}). As higher
order harmonics $\cos(m_q q + m_1\phi_1)$ are generated by the
normalization, there will eventually appear some harmonics causing
important jumps. Recalling that the jumps always take place in the
domain B of Fig.~\ref{fig:melnjump}(a), where the condition
$\dot{q}\approx 2\sqrt{\epsilon}$ should hold, the tuning occurs for a
harmonic satisfying $2m_q\sqrt{\epsilon}+m_1J_{10}\approx 0$.  This
implies a ratio $|m_1|/|m_q|={\cal O}(1/\sqrt{\epsilon})$. In Arnold's
model, such a harmonics will be generated for the first time at the
normalization order $s_0 = |m_1|+|m_q|= {\cal
  O}(1/\sqrt{\epsilon})$. Then, it turns out that there is an optimal
normalization order beyond which the critical harmonic can no longer
be removed from the Hamiltonian. Usual normal form estimates (see
section 4) lead to $s_{opt}={\cal O}(1/\mu^b)$, for a positive
exponent $b$. The size of the harmonic at optimal order will be ${\cal
  O}(\exp(1/\mu^b))$, i.e., i.e., {\it exponentially small} in
$1/\mu$. This, yields, in general, an exponentially small drift
velocity in action space.

An important remark regarding the precise estimates on the speed of Arnold 
diffusion is that the latter depend crucially on whether a system is 
{\it a priori stable} or {\it a priori unstable} (see also section 3 below). 
This distinction has been emphasized in a central paper on the subject 
by~\cite{chiergal-94} (hereafter CG). That paper provides a rigorous 
proof of the occurrence of Arnold diffusion in a priori unstable systems 
and also along the simple resonances of a priori stable systems. It also 
discusses lower bounds on the times necessary for making ${\cal O}(1)$ 
excursions in action space. These bounds are estimated as exponentially 
small in $1/\mu^2$.
\footnote{ Despite the appearances, the paper by CG contains
  several parts accessible to physicists and astrodynamicists. As an
  exercise, readers are invited to study the analogy between several
  rigorous definitions given in CG and the corresponding heuristic
  definitions given in~\cite{chiri-79}, which is addressed to
  physicists. For example, pendulum motions close to the upper and
  lower branches of the pendulum separatrix correspond to the
  `separatrix swings' in CG, the region B where the jumps occur is
  called `origin of the separatrix', the phase sequences $\phi_{1,i}$,
  $i=1,2,...$ of the whisker map are called `phase shifts' (CG section
  4, etc). }

\subsection{Geometric approach}
The arguments exposed so far justify local variations in the values of
the actions $J_1$ and $J_2$, but provide no theory for the long
(${\cal O}(1)$) excursions of the trajectories in the action
space. Demonstration that such excursions are possible requires,
instead, the use of some geometric method. A standard method relies on
the existence of orbits shadowing the heteroclinic intersections
between the stable and unstable invariant manifolds emanating from the
family of hyperbolic tori lying in the phase space of the system under
study.

In Arnold's example, these are the tori $\tau(J_1,J_2)$ defined in
Eq.~(\ref{eq:hyptorus}), which are quite distinct from the
3-dimensional KAM tori referred to subsection 2.1. In particular,
along the tori $\tau(J_1,J_2)$ we always have the invariance
$q(t)=p(t)=0$, corresponding to the hyperbolic fixed point of the
pendulum. However, contrary to what we saw in the previous subsection,
in the geometric method we seek to characterize the motions in the
neighborhood of a hyperbolic torus $\tau(J_1,J_2)$ via the study of
the invariant \textit{asymptotic manifolds} emanating from the torus.

Consider first the case $\mu=0$. We define the stable and unstable
manifolds, ${\cal W}^U_{(0,0)}$, ${\cal W}^S_{(0,0)}$ of the unstable
fixed point of the pendulum as the set of all initial conditions
$(q_0,p_0)$ whose time evolution leads to orbits
$(q(t;q_0,p_0),p(t;q_0,p_0)$ tending asymptotically to the unstable
point $(0,0)$ as $t\rightarrow-\infty$ (for the unstable manifold) or
$t\rightarrow\infty$ (for the stable manifold):
\begin{eqnarray}\label{eq:wsu00}
{\cal W}^U_{(0,0)} &=& \left\{(q_0,p_0)\in\mathbb{T}\times\mathbb{R}: 
\lim_{t\rightarrow-\infty}(q(t;q_0,p_0),p(t;q_0,p_0))=(0,0)\right\} \\
{\cal W}^S_{(0,0)} &=& \left\{(q_0,p_0)\in\mathbb{T}\times\mathbb{R}: 
\lim_{t\rightarrow\infty}(q(t;q_0,p_0),p(t;q_0,p_0))=(0,0)\right\}~~.\nonumber
\end{eqnarray}
For $\mu=0$ the sets ${\cal W}^U_{(0,0)}$, ${\cal W}^S_{(0,0)}$ coincide, as they both 
correspond to the pendulum separatrix. Consider, now, the following set of initial conditions 
of the full problem:
\begin{eqnarray}\label{eq:qini0}
&~&{\cal Q}_0: J_1(0)=J_{10},~~J_2(0)=J_{20},~~\phi_1(0)=\phi_{10},~~\phi_2(0)=\phi_{20}\\
&~&(q(0)=q_0,p(0)=p_0)\in{\cal W}^S_{(0,0)}~~.\nonumber
\end{eqnarray}
Since $\mu=0$ the variables $(q,p)$ evolve independently from the variables $(\phi,J)$. 
Since $(q_0,p_0)\in{\cal W}^S_{(0,0)}$, $(q(t),p(t))$ will tend to $(0,0)$ as 
$t\rightarrow\infty$, while $(\phi,J)$ will have an identical evolution as in the 
torus $\tau(J_1,J_2)$. Hence, the trajectory tends to the torus $\tau(J_{10},J_{20})$ as 
$t\rightarrow\infty$. We then define the stable and unstable manifolds of a torus 
$\tau(J_1,J_2)$ as:
\begin{eqnarray}\label{eq:wsutorus}
{\cal W}^U_{\tau(J_1,J_2)} &=& 
\left\{{\cal Q}_0\in\mathbb{T}^3\times\mathbb{R}^3: 
\lim_{t\rightarrow-\infty} 
\mbox{dist}\left({\cal Q}(t;{\cal Q}_0),\tau(J_1,J_2)\right)=0\right\} \\
{\cal W}^S_{\tau(J_1,J_2)} &=& 
\left\{{\cal Q}_0\in\mathbb{T}^3\times\mathbb{R}^3: 
\lim_{t\rightarrow\infty} 
\mbox{dist}\left({\cal Q}(t;{\cal Q}_0),\tau(J_1,J_2)\right)=0\right\} \nonumber
\end{eqnarray}
where ${\cal Q}(t;{\cal Q}_0)\in\mathbb{T}\times\mathbb{R}^3$ denotes the trajectory 
(in all six variables) corresponding to the initial condition ${\cal Q}_0$. 

\begin{figure}[!ht]
\centering
\includegraphics[width=0.8\textwidth]{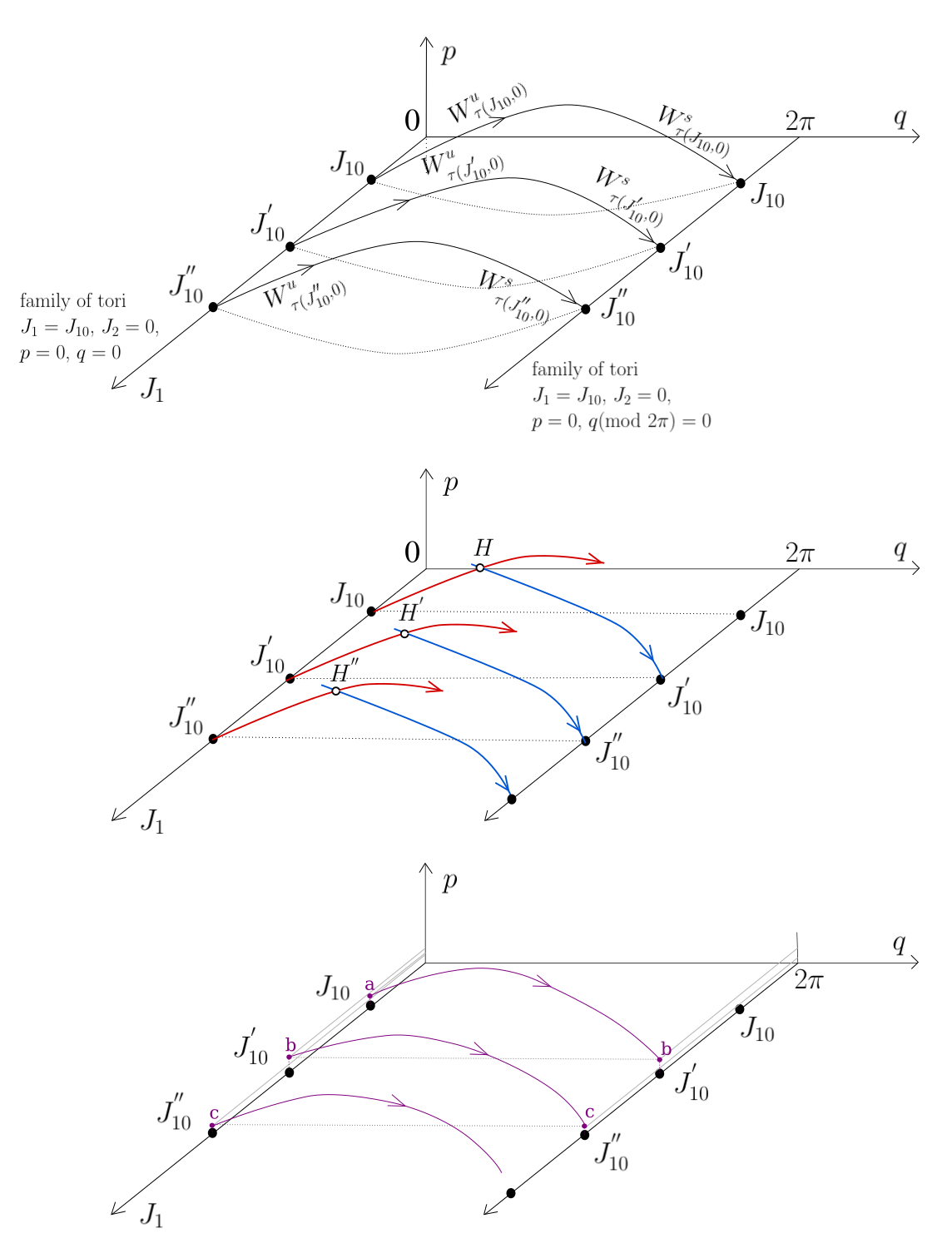}
\caption{Schematic representation of Arnold's mechanism: \textit{Top:}
  When $\mu=0$, the `whiskers' (stable and unstable manifolds) of
  three nearby hyperbolic 2D tori labeled by the actions $J_{10}$,
  $J_{10}'$ and $J_{10}''$ are joined smoothly as pendulum
  separatrices.  \textit{Middle:} For $\mu\neq 0$, the unstable
  manifolds (red) of one torus intersect heteroclinically with the
  stable manifolds (blue) of a nearby torus. This establishes a
  `chain' of heteroclinic connections. \textit{Bottom:} There is a
  true orbit (purple) `shadowing' the above chain, that is, undergoing
  Arnold diffusion.}
  \label{fig:arnsche}
\end{figure}
We saw that the invariant tori $\tau(J_1,J_2)$ (with $q=p=0$) continue
to exist when $\mu\neq 0$. Is it, however, possible to find initial
conditions ${\cal Q}_0$ satisfying the definition of the stable and
unstable manifolds ${\cal W}^S_{\tau(J_1,J_2)}$, ${\cal
  W}^U_{\tau(J_1,J_2)}$ when $\mu\neq 0$? The answer to this question
is affirmative.  In fact, a local normal form around the torus
$\tau(J_1,J_2)$ allows to give in parametric form initial conditions
in the neighborhood of the torus which satisfy the manifold
definition. Then, propagating these local initial conditions backwards
of forwards in time, respectively, we can unfold the whole set of
initial conditions belonging to the manifolds ${\cal
  W}^S_{\tau(J_1,J_2)}$, ${\cal W}^U_{\tau(J_1,J_2)}$ in the perturbed
case as well. However, as argued by Arnold(\cite{Arnold64}; see
also~\cite{chiergal-94}), the manifolds emanating from different tori
in the perturbed system $\mu\neq 0$ have a property not holding when
$\mu=0$, namely, manifolds of tori corresponding to the same energy
but being sufficiently close to each other can {\it intersect
  heteroclinically}, i.e. the unstable manifold of one torus can
interest with the stable manifold of a nearby torus and vice versa.
Figure \ref{fig:arnsche} shows schematically what happens with the
manifolds of the tori $\tau(J_1,J_2)$ in Arnold's model
(\ref{eq:arnold3d}): Consider a fixed value of the energy $E$. On one
such torus we have $q=p=0$, thus $E=J_1^2/2 + J_2$. For every initial
condition with $J_1=J_{1,0}$ we can specify $J_2=E-J_1^2/2$, and thus
define the torus $\tau(J_1=J_{10},J_2=E-J_1^2/2$. In reality, since
$J_2$ is a dummy action variable measuring the change of energy in the
non-autonomous system (\ref{eq:arnold2half}), which is equivalent to
the system (\ref{eq:arnold3d}), only the action $J_{10}$ truly labels
different tori. Hence, for different values of $J_{10}$ we obtain a
family of tori, denoted by $\tau(J_{10},0)$, for different values of
the constant $J_{10}$. The top panel of Fig.~\ref{fig:arnsche} shows
three such tori, $\tau(J_{10},0)$, $\tau(J_{10}',0)$,
$\tau(J_{10}'',0)$, corresponding to three points on the axis $J_1$ of
the figure. In reality, the tori are not points, but they are
parameterized by the angles $\phi_1,\phi_2$ given by all possible
trajectories $\phi_1(t) = \phi_{10}+J_{10}t$, $\phi_2 = t$. These
angular variables are not included in the schematic figure
\ref{fig:arnsche}.

Now, from every torus $\tau(J_{10},0)$ emanate the stable and unstable
manifolds ${\cal W}^S_{\tau(J_{10},0)}$, ${\cal
  W}^U_{\tau(J_{10},0)}$. In the case $\mu=0$, we saw that these
manifolds join each other smoothly, as they actually coincide with the
pendulum separatrix. Hence, as shown in the top panel of
Fig.~\ref{fig:arnsche}, the manifolds of different tori cannot
intersect, i.e., ${\cal W}^U_{\tau(J_{10},0)}$ cannot intersect with
${\cal W}^S_{\tau(J_{10}',0)}$, ${\cal W}^U_{\tau(J_{10}',0)}$ cannot
intersect with ${\cal W}^S_{\tau(J_{10}'',0)}$, etc., no matter how
close the tori $\tau(J_{10},0)$, $\tau(J_{10}',0)$, $\tau(J_{10}'',0)$
are one to the other. However, this changes when $\mu\neq 0$, and it
can be demonstrated that if $\tau(J_{10},0)$ is taken sufficiently
close to $\tau(J_{10}',0)$, the manifolds ${\cal
  W}^U_{\tau(J_{10},0)}$ and ${\cal W}^S_{\tau(J_{10}',0)}$ can
intersect.  The middle panel of Fig.~\ref{fig:arnsche} shows such an
intersection, at the point H, called a heteroclinic point. The
sequence of the heteroclinic points H,H',H'' of the middle panel of
Fig.~\ref{fig:arnsche} will be called a `heteroclinic chain'.  The
sequence of tori whose manifolds yield the points H,H',H'' are known
with various names, namely, the Arnold chain of `whiskered tori' (the
manifolds are the `whiskers'), or the `diffusion path'
(see~\cite{chiergal-94}).

Consider, finally, the past and future trajectories with initial
conditions corresponding to the points H,H',H'', etc. The trajectory
from H belongs to both the invariant manifolds ${\cal
  W}^U_{\tau(J_{10},0)}$ and ${\cal W}^S_{\tau(J_{10}',0)}$.  Thus, in
the limit $t\rightarrow\-\infty$ the trajectory tends to the torus
$\tau(J_{10},0$, while, in the limit $t\rightarrow\infty$ the
trajectory tends to the torus $\tau(J_{10}',0$. This implies that this
particular trajectory undergoes no large excursion in the action
space, since its past and future is confined between two nearby
asymptotic limits. Similarly, the past and future from the
heteroclinic point H' connect the tori $\tau(J_{10}',0)$ with
$\tau(J_{10}'',0)$, those from the heteroclinic point H'' connect the
tori $\tau(J_{10}'',0)$ with $\tau(J_{10}''',0)$, etc., but the
corresponding trajectories make only bounded excursions in the action
space. However, employing a so-called {\it shadowing lemma}, it is
possible to demonstrate that there is one continuous in time
trajectory of the system which remains piece-wise close
(i.e. `shadows') any one of the distinct heteroclinic trajectories
from the points $H$,$H'$,$H''$,... Such a trajectory is shown
schematically in the last panel of Fig.~\ref{fig:arnsche}. It is
precisely this trajectory which materializes the `Arnold's mechanism'
referred to in the introduction. Extending the heteroclinic chain
$H,H',H'',...,H^{(n)},...$ to include more heteroclinic points, one
can find a trajectory connecting the neighborhoods of the initial torus
$\tau(J_{10},0)$ and another torus $\tau(J_{10}^{(n)},0)$ located at
arbitrarily large distance from $\tau(J_{10},0)$ (possibly limited
only by the requirement of the two tori being isoenergetic).

\begin{figure}[!ht]
  \centering
  \includegraphics[width=1.\textwidth]{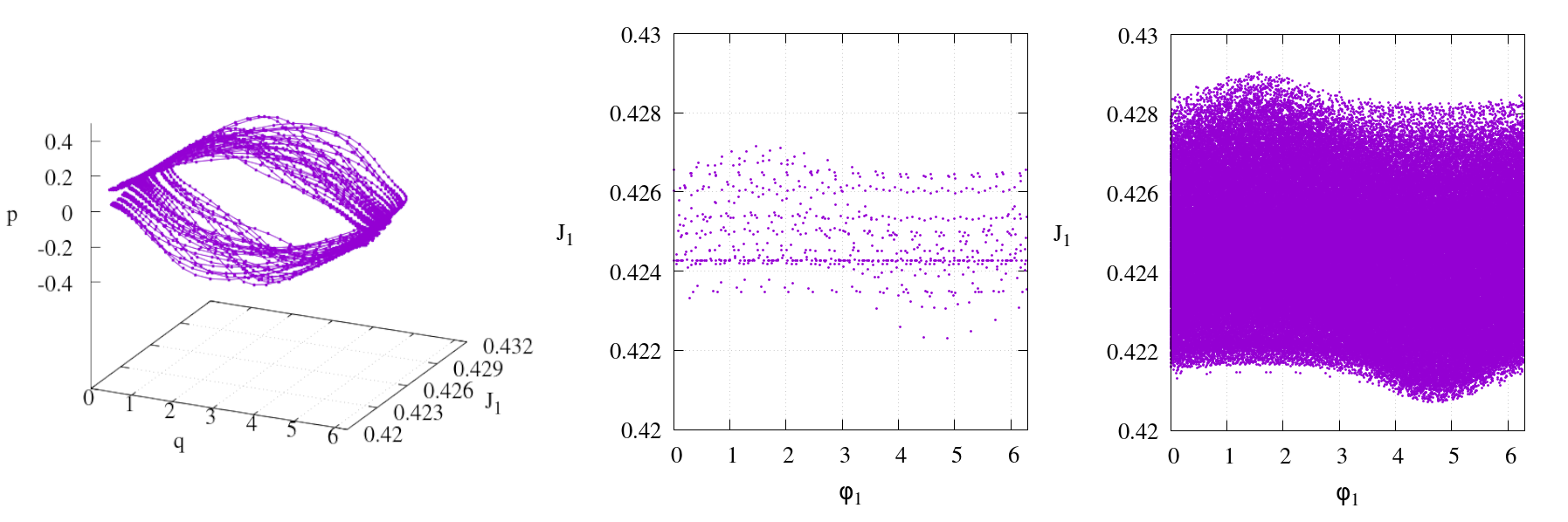}
  \caption{\textit{Left:} Real (non-schematic) orbit shadowing the
    intersecting manifolds of nearby tori in Arnold's model, obtained
    by plotting in the $(p,q) \times J_1$ space the same orbit as in
    Fig.~\ref{fig:arnoldex}, for the integration time
    $t=700$. \textit{Center} and \textit{Right:} the projection of
    the orbit on the $(\phi_1,J_1)$ plane at two different integration
    times, $t=700$ and $t=150000$.}
  \label{fig:arnoldex2}
\end{figure}
Does the `Arnold mechanism' interpret the long-term evolution of the
numerical trajectory used in our example in the previous subsection?
Figure \ref{fig:arnoldex2} suggests this to be so, provided that the
trajectory is integrated for times much longer than those referred to
in the previous subsection. The left panel shows how the trajectory
produced by integration of the complete model (\ref{eq:arnold3d}), and
with the same initial conditions as in Fig.~\ref{fig:arnoldex} shadows
the whiskers of nearby tori $\tau(J_1,J_2)$. The middle and right panels 
show the projection of the trajectory in the plane $(\phi_1,J_1$. Clearly,
the trajectory remains piece-wise close to various rotational tori
(corresponding to different values of $J_1$), however, as the
integration time extends from $t=700$ to $t=1.5\times 10^5$ the
excursion in $J_1$ extends from a total size $\sim 10^{-2}$ to nearly
$\sim 10^{-1}$. Note that as the trajectory reaches domains further
and further away from this particularly selected initial condition,
the drift in action space actually gets slower (see last paragraph of
subsection 2.2).

As a final remark, the above geometric picture of intersecting
manifolds can be extended, from the chain of nearby tori, to include
the whole invariant set ${\cal T}$ of the tori $\tau(J_1,J_2)$. This
is a four-dimensional subset of $\mathbb{R}^2\times\mathbb{T}^2$,
which is \textit{normally hyperbolic} (see \cite{delshetal-08b} for
definitions). Normal hyperbolicity implies the existence of a stable
and unstable manifold for the whole invariant set $\cal{T}$. Since, in
Arnold's example, $\cal{T}$ is just foliated by the tori
$\tau{J_1,J_2}$, the manifolds ${\cal W}^U_{\cal T}$, ${\cal
  W}^S_{\cal T}$ are just the union of the unstable and stable
manifolds of all the tori. Homoclinic orbits can then be described by
a `scattering map' indicating how a point on ${\cal T}$ is mapped
asymptotically in time to another point on ${\cal T}$ via a
doubly-asymptotic orbit.

\section{A priori stable systems - Nekhoroshev theory}
Consider the following Hamiltonian in action-angle variables, which,
according to Poincar\'{e}~\cite{poincare}, represents the {\it
  ``fundamental problem of dynamics''}:
\begin{equation}\label{eq:hamgen}
H(\phi,I) = H_0(I)+\epsilon H_1(\phi,I)
\end{equation}
with $\phi\in\mathbb{T}^n$, $I\in\mathbb R^n$. 

For $\epsilon=0$ the system is integrable $H=H_0(I)$ and the phase space is foliated by 
invariant tori labeled by the constant actions $I$. On each torus the angles evolve 
linearly with the frequencies $\omega(I)=\nabla_IH_0(I)$. Periodic orbits, or, in general, 
tori of dimension $n'<n$ correspond to values of the actions $I$ for which the frequencies 
$\omega(I)$ satisfy $n-n'$ commensurability conditions. However, all these low-dimensional 
objects are neutral in stability, and there are no separatrices or any other type of 
asymptotic manifolds (`whiskers') associated to them. In other words, there is no in-built 
hyperbolicity in the Hamiltonian $H_0(I)$. Hence, invariant objects of (partially) 
hyperbolic character can only be born by setting $\epsilon\neq 0$. Such systems were 
thus called (by CG) `a priori stable'.  

The lack of invariant phase space objects with inherent hyperbolicity generates several 
challenging new questions regarding Arnold diffusion. We now summarize some of these 
questions as well as known results related to Arnold diffusion in a priori stable systems. 

\subsection{Nekhoroshev theory and exponential stability}
Whatever the mechanism possible to cause Arnold diffusion in an a
priori stable system, the speed of the drift in action space in such a
system is bounded before all by the {\it Nekhoroshev theorem}
(\cite{nekho-77}, \cite{benetal-85}, \cite{bengall86},\cite{Lochak-92},
 \cite{poshel-93}): \\ \\
{\bf Nekhoroshev theorem:} {\it Assume a
  Hamiltonian of the form (\ref{eq:hamgen}) with $\epsilon>0$, with
  $H$ analytic in a complex extension ${\cal D}$ of the set
  $D\times\mathbb{T}^n$, where $D\subset\mathbb{R}^n$ is open, and
  $H_1$ bounded.  Assume that $H_0$ satisfies suitable {\it steepness}
  conditions. Then, there are positive constants $a,b,\epsilon_0$ such
  that, for $\epsilon<\epsilon_0$ and for all initial conditions in
  ${\cal D}$, under the flow of the Hamiltonian $H$ we have:}
\begin{equation}\label{eq:nekho}
|J(t)-J(0)|<\epsilon^a~~~\mbox{for all times~$t<T_N$ with~}
T_N={\cal O}\left({\epsilon_0\over\epsilon}\exp((\epsilon_0/\epsilon)^b)\right)
\end{equation}
We refer to $T_N$ as the `Nekhoroshev time'. 
A detailed discussion of the meaning and importance of `steepness' in the above theorem 
is made in~\cite{guzzetal-11}\cite{SchiGuz-13}\cite{chieguzzo-19}. 
We briefly refer to steepness in subsection 3.2 below. 

Demonstration of the Nekhoroshev Theorem (see~\cite{Pisa} for a
tutorial) requires combining an {\it analytical} with a {\it
  geometric} part. The analytical part deals with the local
construction of a `Nekhoroshev normal form', whose remainder at the
optimal normalization order turns to be exponentially small. On the
other hand, the geometric part deals with the construction of a set of
subdomains $D_1,D_2,\ldots\subset{\cal D}$ defined so that: 
i) a different local normal form with exponentially small remainder can
be constructed in each domain, and ii) the union of all domains
provides a covering of ${\cal D}$. The structure of {\it resonant
  manifolds} (see below), depending on the form of the integrable part
$H_0(I)$ of the Hamiltonian, as well as the size of the analyticity
domain around each manifold, determined by the form of $H_1(\phi,I)$,
are crucial factors in the appropriate definition of the domains
$D_i$. In particular, the domains $D_i$ must have size depending
algebraically on $\epsilon$, i.e. $\mbox{diam}(D_i)={\cal
  O}(\epsilon^{a_i})$, $a_i>0$. One then demonstrates that this
dependence allows to obtain a covering of ${\cal D}$ by combining many
such domains when $\epsilon$ is arbitrarily small
(see~\cite{morbiguz-97} for a heuristic argument).  Now, the size of
the optimal remainder of each local normal form scales as
$||R||=O\left(-\exp((\epsilon_{0,i}/\epsilon)^{b_i})\right)$, for some
positive constant $\epsilon_{0,i}$ and positive exponent
$b_i$. Choosing the worst possible combination $a_i,b_i$ and
$\epsilon_{i,0}$ from those holding in each domain allows to arrive at
the global bound (\ref{eq:nekho}). In practice, locally we can obtain
better bounds using the local parameters $a_i,b_i,\epsilon_{0,i}$. It
turns out that the exponents $a,b$ depend on i) the number of degrees
of freedom $n$, ii) the so-called steepness indices holding within the
domain (see \cite{guzzetal-11} for definitions) and, finally, iii) the
\textit{multiplicity} of the local resonance considered (see below).

It is noteworthy that, while in the proof of the theorem the analytical part 
plays a minimal role, the {\it actual construction} of the Nekhoroshev normal 
form in any explicit application implies reaching a very high order of 
normalization, involving typically millions of operations that can only be 
carried out with the aid of a computer-algebraic program. Starting from the 
sixties  (\cite{conto-60},\cite{contomout65}, \cite{gustavson},
\cite{gio-79}), such programs dealt first with the simpler case of systems
with elliptic equilibria, such as the celebrated H\'{e}non-Heiles
system~\cite{hehe-64}. In such systems, exponential estimates can be obtained 
without the need of a geometric construction as the one of the Nekhoroshev 
theorem. Well known applications in Celestial Mechanics have been given, 
referring, for example, to the long term stability of the Trojan asteroids 
of Jupiter \cite{celgio-91}\cite{giosko-97}\cite{eftsan-05}\cite{lhoetal-08}, 
the spin-orbit problem \cite{sansoetal-14}, and the $J_2$ problem 
of satellite motions \cite{steigio-97}\cite{deblaetal-21}. On the other hand, 
computing the optimal Nekhoroshev normal form in a generic Hamiltonian of the 
form (\ref{eq:hamgen}) has been possible so far only in simple models with
$n=3$ degrees of freedom \cite{efthy-08}\cite{efthyhar-13}\cite{cinetal-14}
\cite{GEP}. Such computations allow for a direct
comparison between `semi-analytical' (i.e. by the remainder of the
Nekhoroshev normal form) and numerical results on the speed of Arnold
diffusion, as well as on the adiabatic evolution of the action
variables in a priori stable systems. Most notable among the numerical
experiments are those carried over the years by the group of
C. Froeschl\'{e}, M. Guzzo and E. Lega (\cite{legaetal-03}\cite{guzzoetal-05}
\cite{Frosetal-05}\cite{guzzetal-11}), which have given clear
evidence of the occurrence of Arnold diffusion in a priori stable
systems. A comparison of the exponents $a,b$ found by the Nekhoroshev
normal form construction and by the numerical experiments has shown a
very good agreement. This has extended also to estimates on the
coefficient of Arnold diffusion as well as to the modeling of the
jumps carried by the adiabatic action variables along the heteroclinic
transitions taking place in single resonance domains. In the sequel we
give a summary of the above results with the help (as in the previous
section) of a simple example of a priori stable system with $n=3$
degrees of freedom.

\subsection{A simple example}
Consider the 3DOF Hamiltonian in action-angle variables:
\begin{equation}\label{eq:hamfr}
  H = H_0+\epsilon H_1 = \frac{I_1^2}{2} - \frac{I_2^2}{2}+ \frac{I_2^3}{3 \pi}
  +  2 \pi I_3 + {\epsilon\over
4+\cos\phi_1+\cos\phi_2+\cos\phi_3}~~.
\end{equation}
The Hamiltonian (\ref{eq:hamfr}) has been used in~\cite{GEP} in the   
study of the evolution of the adiabatic action variables. An analogous 4D symplectic 
mapping was used in \cite{guzzetal-11} for the study of the effects of steepness on 
the stability of the orbits. 

The flow corresponding to the integrable part of (\ref{eq:hamfr})
\begin{equation}\label{eq:hamfr0}
  H_0 = \frac{I_1^2}{2} - \frac{I_2^2}{2}+ \frac{I_2^3}{3 \pi} +  2 \pi I_3~~.
\end{equation}
is given by $\dot{I_i}=0$, $i=1,2,3$ and $\dot{\phi}_1=\omega_{0,1}=I_1$, 
$\dot{\phi}_2=\omega_{0,2}=-I_2 +\frac{1}{\pi} I_2^2$, $\dot{\phi}_3=\omega_{0,3}=2\pi$. 
Thus, all trajectories lie on invariant tori labeled by the actions $I_i$ or the 
corresponding frequencies $\omega_{0,i}$. 

Let $k\equiv(k_1,k_2,k_3)\in \mathbb{Z}^3$. We call {\it resonant manifold} 
${\cal RM}(k_1,k_2,k_3)$ associated to the Hamiltonian $H_0$ the two-dimensional manifold
\begin{eqnarray}\label{eq:resman}
{\cal RM}(k_1,k_2,k_3)&:=
&\bigg\{(I_1,I_2,I_3)\in\mathbb{R}^3: \\
&~&k\cdot\omega_0(I)= 
k_1I_1+k_2(-I_2 +\frac{1}{\pi} I_2^2)+k_3 \,2 \pi=0\bigg\}~~.\nonumber
\end{eqnarray}
We call \textit{energy manifold} ${\cal E}(E)$ the two-dimensional manifold 
\begin{equation}\label{eq:eneman}
{\cal E}(E):=
\left\{(I_1,I_2,I_3)\in\mathbb{R}^3:~~ 
H_0(I)={1\over 2}(I_1^2-I_2^2)+\frac{I_2^3}{3 \pi}+2\pi I_3=E\right\}~~.
\end{equation}

\begin{figure}[h]
\centering
\includegraphics[width=1.\textwidth]{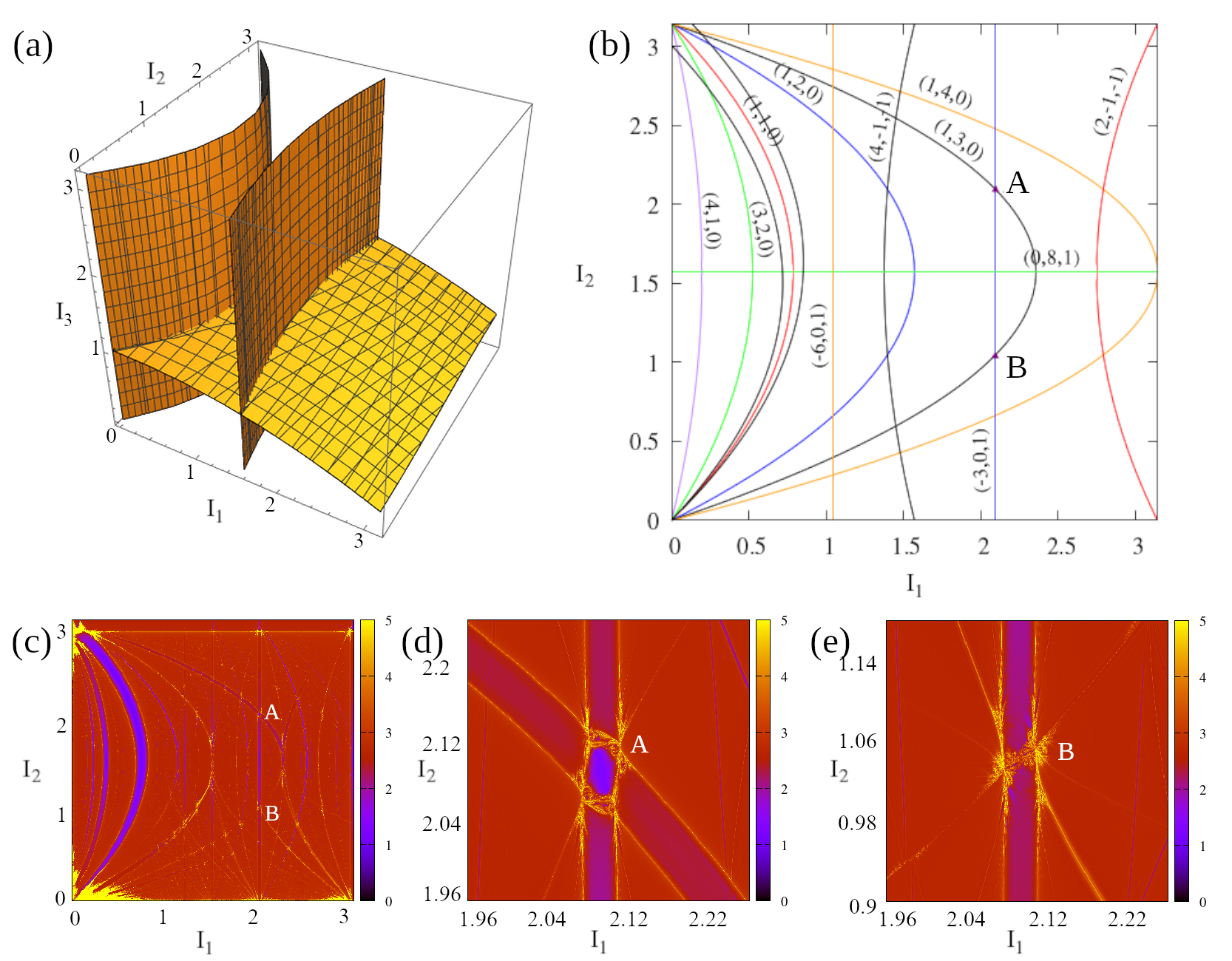}
\caption{(a) Part of the energy manifold ${\cal E}(E)$ in the model
  (\ref{eq:hamfr0}) for $E=1$ (yellow), intersected by parts of the
  resonant manifolds ${\cal RM}(1,1,0)$ and ${\cal RM}(4,-1,-1)$
  (orange).  (b) Projection of the Arnold web of resonances on the
  $(I_1,I_2)$ plane. For the resonance (1,1,0) the corresponding
  separatrix borders are also displayed as computed theoretically for
  $\epsilon = 0.05$ (see text). (c) FLI stability map for the
  Hamiltonian (\ref{eq:hamfr}) with $\epsilon = 0.05$. The web of
  resonances is visualized through the detection of weakly chaotic
  orbits at the borders of each resonance. (d) and (e) Details of
  figure (c) around the resonance junctions A and B, respectively,
  where the resonant manifolds ${\cal RM}(1,3,0)$ and ${\cal
    RM}(3,0,-1)$ intersect.}
\label{fig:arnweb}
\end{figure}

Figure \ref{fig:arnweb}(a) shows a part of the energy manifold ${\cal
  E}(E)$ for $E=1$ as well as parts of the two resonant manifolds
${\cal RM}(1,1,0)$ and ${\cal RM}(4,-1,-1)$.  The set of all curves
formed by the intersection of all resonant manifolds ${\cal RM}(k)$,
$k\in\mathbb{Z}^3$, $|k|\neq 0$ with the energy manifold ${\cal E}(E)$
is called the {\it Arnold web} (or `web of resonances'). In our
example, the definition of the resonant manifolds via
Eq.~(\ref{eq:resman}) does not depend on $I_3$. Thus all resonant
manifolds intersect normally the plane $(I_1,I_2)$ at curves given by
Eq.~(\ref{eq:resman}).  Figure~\ref{fig:arnweb}(b) shows some of these
resonant curves marked with the corresponding integers
$(k_1,k_2,k_3)$.

The set of the resonant curves defined by all possible
$(k_1,k_2,k_3)\in\mathbb{Z}^3$, $|k|\neq 0$ is dense in the square
$S(I_1,I_2)$ depicted in Fig.~\ref{fig:arnweb}(b): for any open, small
whatsoever, neighborhood $S_i\subset S(I_1,I_2)$ there exist
integers$(k_1,k_2,k_3)$ such that the corresponding resonant curve
crosses $S_i$. However, not all these resonances are equally important
for dynamics. This is evidenced by computing a \textit{stability map}
in the same square $S(I_1,I_2)$ via the use of a chaotic
indicator. Figure~\ref{fig:arnweb}(c) shows the stability map computed
by the Fast Lyapunov Indicator (FLI,~\cite{Frosetal-97} and the chapter 
by Guzzo and Lega in this book) in a grid of
initial conditions for $(I_1,I_2)$, setting initially
$I_3=\phi_1=\phi_2=\phi_3=0$, and for an integration time $t=1000$. We
immediately note that the FLI map in Fig.~\ref{fig:arnweb}(c) is able
to depict the structure of the Arnold web in great detail. This fact,
first found in~\cite{Frosetal-00} has played a crucial role in the
numerical study of Arnold diffusion in a priori stable systems.

In Fig.~\ref{fig:arnweb}(c) we see that the most prominent structures
are related to low order resonances ($|k|=|k_1|+|k_2|+|k_3|$
small). Also, we notice that, for some resonances (e.g.  (1,1,0)), the
FLI map shows a double set of curves going nearly parallel one to the
other along the resonance, with a blue zone between the curves. Other
resonances, instead, are identified by a single line (yellow). This
distinction depends on the sign of the Fourier coefficient of the
corresponding resonant harmonics in the function $H_1$ of
Eq.~(\ref{eq:hamfr}).  We have:
$$
{1\over 4+\cos\phi_1+\cos\phi_2+\cos\phi_3} = 
\sum_{k_1,k_2,k_3=-\infty}^{\infty}
h_{k_1,k_2,k_3}\cos(k_1\phi_1+k_2\cos\phi_2+k_3\phi_3)
$$
where $h_{k_1,k_2,k_3}$ can be easily computed expanding the denominator in Taylor series and 
using the trigonometric reduction formulas. Consider a toy Hamiltonian in which only one 
harmonic is isolated:
\begin{equation}\label{eq:hamrestoy}
  H_{res} = \frac{I_1^2}{2} - \frac{I_2^2}{2}+ \frac{I_2^3}{3 \pi}
  +  2 \pi I_3 + \epsilon h_{k_1,k_2,k_3}\cos(k_1\phi_1+k_2\cos\phi_2+k_3\phi_3)~~.
\end{equation}
Such a model will be obtained by just performing one step of
perturbation theory eliminating from the Hamiltonian (\ref{eq:hamfr})
all other harmonics except for the resonant one (see section 4). Now,
the Hamiltonian (\ref{eq:hamrestoy}) is integrable. To show this,
assume (without loss of generality) $k_1\neq 0$.  Consider two
linearly independent integer vectors $m,n\in\mathbb{Z}^3$ such that
$m\cdot k=n\cdot k=0$ (for example $m=(k_2,-k_1,0)$, $n=(k_3,0,-k_1)$).
Consider the canonical transformation
$(\phi_1,\phi_2,\phi_3,I_1,I_2,I_3)$ $\rightarrow$
$(\phi_R,\phi_{F1},\phi_{F2},I_R,I_{F1},I_{F2})$ defined by
$$
\phi_{R}=k\cdot\phi, \phi_{F1}=m\cdot\phi, \phi_{F2}=n\cdot\phi~~,
$$
as well as the inverse of the equations
\begin{eqnarray}\label{eq:restra}
I_1&=&k_1I_R+m_1I_{F1}+n_1I_{F2},  \nonumber\\
I_2&=&k_2I_R+m_2I_{F1}+n_2I_{F2},  \\
I_3&=&k_3I_R+m_3I_{F1}+n_3I_{F2}~~.\nonumber
\end{eqnarray}
Substituting these expressions into (\ref{eq:hamrestoy}) we arrive at:
\begin{equation}\label{eq:hamrestoy2}
H_{res}= H_0(I_R,I_{F1},I_{F2}) + \epsilon h_{k_1,k_2,k_3}\cos(\phi_R)~~.
\end{equation}
Since the angles $\phi_{F1},\phi_{F2}$ are ignorable, the above model has two integrals 
of motion $I_{F1},I_{F2}$ besides the energy. We are interested in studying the behavior 
of the model $H_{res}$ in a neighborhood around values $(I_{1*},I_{2*},I_{3*})$ which 
satisfy the resonance exactly. Setting $I_i = I_{i*}+J_i$, $i=1,2,3$ and substituting 
into (\ref{eq:hamrestoy}) we arrive at:
\begin{eqnarray}\label{eq:h0exp}
H_0(J) &=& 
H_0(I_*) + \nabla_IH_0(I_*)\cdot J 
+ {1\over 2}
\sum_{i=1}^3\sum_{j=1}^3 {\partial^2 H_0(I_*)\over\partial I_i\partial I_j}J_iJ_j \\
&+& {1\over 6}\sum_{i=1}^3\sum_{j=1}^3\sum_{l=1}^3 
{\partial^3 H_0(I_*)\over\partial I_i\partial I_j\partial I_l}J_iJ_jJ_l+\ldots~~. \nonumber
\end{eqnarray}
The constant term $H_0(I_*)$ can be omitted. The term $\nabla_IH_0(I_*)\cdot J$ has the form
$$
\nabla_IH_0(I_*)\cdot J = 
(k\cdot\omega_*)J_R +(m\cdot\omega_*)J_{F1}+(n\cdot\omega_*)J_{F2} = 
(m\cdot\omega_*)J_{F1}+(n\cdot\omega_*)J_{F2}~~.
$$
where $\omega_*$ denotes the vector of the resonant frequencies $\omega_{i*}=\omega_i(I_*)$, 
and the variables $J_R, J_{F1}, J_{F2}$ are defined as $J_R = I_R-I_{R*}$, 
$J_{F1} = I_{F1}-I_{F1*}$, $J_{F2} = I_{F2}-I_{F2*}$ with 
$$
\left(
\begin{array}{c}
I_{R*}\\
I_{F1*}\\
I_{F2*}
\end{array}
\right)
=
\left(
\begin{array}{ccc}
k_1 &m_1 &n_1\\
k_2 &m_2 &n_2\\
k_3 &m_3 &n_3
\end{array}
\right)
\left(
\begin{array}{c}
I_{1*}\\
I_{2*}\\
I_{3*}
\end{array}
\right)
$$
The frequencies $\omega_*$ satisfy $k\cdot\omega_*=0$, hence the transformed Hamiltonian 
contains linear terms only 
for the `fast' action variables $J_{F1},J_{F2}$. Instead, the resonant action $J_R$ appears 
in the Hamiltonian only in quadratic terms (or of higher degree) in the actions. Setting the 
integrals as $J_{F1}=0$, $J_{F2}=0$ implies the relations $I_{F1} = I_{F1*}$, 
$I_{F2}=I_{F2*}$, that is:
\begin{eqnarray}\label{eq:plfdmot}
I_1&=&k_1I_R+m_1I_{F1*}+n_1I_{F2*}\nonumber\\
I_2&=&k_2I_R+m_2I_{F1*}+n_2I_{F2*}\\ 
I_3&=&k_3I_R+m_3I_{F1*}+n_3I_{F2*}\nonumber
\end{eqnarray}
Thus, the motion in all three action variables under the flow of the
model Hamiltonian (\ref{eq:hamrestoy}) is determined by the only
evolving action, namely $I_R$, and it is confined along a line ${\cal
  L}(I_*)$ in the space $(I_1,I_2,I_3)$ defined parametrically by
Eq.~(\ref{eq:plfdmot}). The projection of the line ${\cal L}(I_*)$ on
the plane $(I_1,I_2)$ is given by
\begin{equation}\label{eq:plfd}
I_2 = {1\over k_1}\left(-k_2I_I+(k_2m_1-k_1m_2)I_{F1*}+(k_2n_1-k_1n_2)I_{F2*}\right) 
\end{equation}
Also, the only non-ignorable angle in the model Hamiltonian of the resonance is 
$\phi_R\in\mathbb{T}$. The set ${\cal P}_F(I_*)={\cal L}(I_*)\times\mathbb{T}$ is 
called {\it plane of fast drift}. On this plane the motion is described by a 
pendulum-like Hamiltonian in the local variables $(\phi_R,J_R)$. The equations 
(\ref{eq:plfdmot}) imply $J_R = (k\cdot J)/(k\cdot k)$. Then, the quadratic term  
in the actions in (\ref{eq:h0exp}) takes the form:
\begin{equation}\label{eq:beta}
{1\over 2}
\sum_{i=1}^3\sum_{j=1}^3 {\partial^2 H_0(I_*)\over\partial I_i\partial I_j}J_iJ_j
={1\over 2}\beta(I_*)J_R^2~~\mbox{with}~~\beta(I_*)={1\over k^2}(M(I_*)k)\cdot k
\end{equation}
where $M(I_*)$ is the $3\times 3$ Hessian of the Hamiltonian $H_0$ calculated at the 
point $I_*$
$$
M_{ij}(I_*) = \left({\partial^2H_0\over\partial I_i\partial I_j}\right)_{I=I_*}
$$
Similarly, the cubic term in the actions takes the form $(1/3)\gamma(I_*)J_R^3$ with
\begin{equation}\label{eq:gamma}
\gamma(I_*)={1\over 2|k|^{3/2}}
\sum_{i=1}^3\sum_{j=1}^3\sum_{l=1}^3 
\left({\partial^3 H_0(I_*)\over\partial I_i\partial I_j\partial I_l}\right)_{I=I_*}
k_ik_jk_l
\end{equation}
Hence, apart from constants we have
\begin{equation}\label{eq:hrespend}
H_{res}={1\over 2}\beta(I_*)J_R^2 + {1\over 3}\gamma(I_*)J_R^3 + \epsilon h_k\cos(\phi_R)
\end{equation}
where, in the model (\ref{eq:hamfr0}) we get:
\begin{equation}\label{eq:betgam}
\beta(I_*) =k_1^2+k_2^2\left({2I_{2*}\over\pi}-1\right),~~~\gamma(I_*)={k_2^3\over\pi}
\end{equation}
Except for the case $k_1=k_2$ and $I_{2*}\rightarrow 0$, the coefficient $\beta(I_*)$ 
is in general a ${\cal O}(1)$ quantity. Then, taking $J_R$ in a domain of size 
${\cal O}(\epsilon^{1/2})$, the  term ${1\over 2}\beta(I_*)J_R^2$ is more important than 
the term ${1\over 3}\gamma(I_*)J_R^3$ in $H_{res}$. This means that $H_{res}$ (ignoring 
cubic terms) becomes a pendulum Hamiltonian with separatrices extending in a domain  
$J_{R,min}\leq J_R\leq J_{R,max}$ estimated by:
\begin{equation}\label{eq:sepwidth}
J_{R,min}\simeq -2\left({\epsilon\over|\beta(I_*)|}\right)^{1/2},~~~ 
J_{R,max}\simeq 2\left({\epsilon\over|\beta(I_*)|}\right)^{1/2}~~. 
\end{equation}
In reality, the motion very close to the separatrix will be weakly
chaotic, due to the fact that, as discussed below, the remaining
resonances can be eliminated only up to an exponentially small
remainder, and hence there is some degree of chaos due to the
interaction of these resonances with the principal one
$(k_1,k_2,k_3)$. The motion along the separatrix-like thin chaotic
layer of the resonance can be projected also on the plane
$(I_1,I_2)$. The projection is constrained in a segment along the
line ${\cal L}(I_*)$, which represents the intersection of the plane
of fast drift with the plane $(I_1,I_2)$. In particular, the motion
along the separatrix layer projects to a linear segment given by
Eq.~(\ref{eq:plfdmot}), setting $I_R = I_{R*}+J_R$, and varying $J_R$
in the limits $J_{R,min}\leq J_R\leq J_{R,max}$.

We are now able to understand the structure of the FLI map shown in
Fig.~\ref{fig:arnweb}(c).  Let $I_*$ be one point along the resonance
$(k_1,k_2,k_3)$. Since in the computation of the FLI we have set the
initial conditions $\phi_i=0$, $i=1,2,3$, the FLI map intersects the
plane of fast drift crossing the point $I_*$ at the value
$\phi_R=0$. Whenever the coefficients $\beta(I_*)$ and
$h_{k_1,k_2,k_3}$ have the same sign, the point $\phi_R$ represents
the unstable equilibrium point of the Hamiltonian $H_{res}$. One has
$J_R=0$ there, thus, by Eqs.~(\ref{eq:plfdmot}) we get a unique point
on the FLI map, given by $I_1=I_{1*}$, $I_2=I_{2*}$. On the contrary,
when $\beta(I_*)$ and $h_{k_1,k_2,k_3}$ have opposite signs, the point
$\phi_R$ corresponds to the stable equilibrium point of the
Hamiltonian $H_{res}$. Then, the line $\phi_R=0$ on the fast drift
plane crosses the separatrix layer approximately at the values
$J_R=J_{R,min}$ and $J_R=J_{R,max}$.  Thus, by Eqs.~(\ref{eq:plfdmot})
we get two point on the FLI map, given by $I_1=I_{1*}+k_1J_{R,min}$,
$I_2=I_{2*}+k_2J_{R,min}$, and $I_1=I_{1*}+k_1J_{R,max}$,
$I_2=I_{2*}+k_2J_{R,max}$. Joining the two families of points
representing the separatrices for different points $I_*$ along the
same resonance yields two curves on the plane $(I_1,I_2)$ which follow
nearly parallelly the curve of the resonance, having between
themselves a ${\cal O}(\epsilon^{1/2})$ distance. Figure
\ref{fig:arnweb} shows the two curves marking the borders of the
resonance (1,1,0), as computed by the above formulas. This fits very
well the borders found by the FLI map of Fig.~\ref{fig:arnweb}(c). The
blue zone between the two borders corresponds to regular orbits, which
are the libration orbits of the pendulum for initial conditions inside
the separatrices.

In general, fixing a certain model $H_0$, we have
$\mbox{sign}[\beta(I_*)]= \mbox{sign}[(M(I_*)k)\cdot k]$. When the
quadratic form $(M(I_*)k)\cdot k$ is positive definite, $\beta(I_*)$
has always the same sign, independently of the resonant vector $k$.
In this case, whether the separatrices intersect with the chosen
section at a single or double curve depends only on the sign of the
coefficient $h_k$ of the Fourier harmonic $\cos(k\cdot\phi)$ in
$H_1$. On the contrary, if the Hessian matrix $M(I_*)$ is not positive
definite, the sign of $\beta(I_*)$ depends on the value of $I_*$ and
on the choice of resonance, i.e., of the vector $k$. In the model
(\ref{eq:hamfr0}), we readily find that $M(I_*)$ is positive definite
in the semi-plane $I_{2*}>\pi/2$, while it is not in the semi-plane
$I_{2*}<\pi/2$. In the latter one, the sign of $\beta$ depends on the
particular choice of resonance. For example, for the resonance
$k=(1,1,0)$ there is no change of sign of $\beta(I_*)$ across the two
semi-planes. For all other resonances $k=(1,k_2,0)$, $k_2>1$,
$\beta(I_*)$ changes sign, instead, at the value
$I_{2*}=(\pi/2)(1-k_2^2/k_1^2)$, a fact easily verified by carefully
inspecting the FLI map of Fig.(\ref{fig:arnweb}).

Besides graphical consequences for the FLI maps, positive-definiteness
(or not) of the Hessian matrix $M(I_*)$ affects several aspects of the
dynamics: an important aspect regards the dynamics around
\textit{resonance junctions}. In the case with $n=3$ DOF, we consider
points $I_*$ for which there exist two linearly independent non-zero
integer vectors $k^{(1)}$, $k^{(2)}$ satisfying:
\begin{equation}\label{eq:dbleres}
k^{(1)}\cdot\omega(I_*)=0,~~~ k^{(2)}\cdot\omega(I_*)=0~~.
\end{equation}
Such points $I_*$ are said to belong to resonant junctions of
multiplicity 2: this is a curve, in the 3D action space, where all
resonant manifolds ${\cal RM}(\lambda_1k^{(1)}+ \lambda_2k^{(2)})$
defined by the two linearly independent vectors $k^{(1)},k^{(2)}$ and
by $\lambda_1,\lambda_2\in\mathbb{Z}$ intersect each other. For $n=3$
a resonant junction can only be of multiplicity 2. For $n>3$, instead,
resonance junctions can be of multiplicity $2\leq mult\leq n-1$), and
the corresponding resonant junctions are manifolds of dimension
$n-mult$.

Figures~\ref{fig:arnweb}(d) and (e) show the FLI maps around the resonance
junctions formed by the crossing of the resonances $(1,3,0)$ and
(3,0,-1) at the points A and B. We immediately notice the difference
in structure of the resonance crossings at these two points. Briefly,
this can be understood as follows (see~\cite{LaPlata} for details):
let $I_*$ be a doubly resonant point. Define the vector
$m=k^{(1)}\times k^{(2)}$ as well as the canonical transformation:
\begin{eqnarray}\label{eq:trdble}
J_i&=&k_i^{(1)}J_{R1}+k_i^{(2)}J_{R2}+m_iJ_F,~~~i=1,2,3 \nonumber\\
\phi_{R1}&=&k^{(1)}\cdot\phi,~~~\phi_{R2}=k^{(2)}\cdot\phi,~~~
\phi_{F}=m\cdot\phi
\end{eqnarray}
where, as before, $J_i=I_i-I_{i*}$. By Eq.(\ref{eq:h0exp}) up to
quadratic terms we now get (apart from a constant)
\begin{eqnarray}\label{eq:h0dblequad}
  H_0 &=& \omega_F J_F \\
    &+& {1\over 2}\sum_{i=1}^3\sum_{j=1}^3 M_{ij}(I_*)
(k_i^{(1)}J_{R1}+k_i^{(2)}J_{R2}+m_iJ_F)(k_j^{(1)}J_{R1}+k_j^{(2)}J_{R2}+m_jJ_F)
\nonumber
\end{eqnarray}
The frequency $\omega_F = m\cdot\omega$ yields the rate of change of
the unique `fast angle' of the problem $\phi_F=m\cdot\phi$ (conjugate
to $J_F$). As before, we can assume computing a resonant normal form
which eliminates all harmonics in the problem except
$\cos((\lambda_1k^{1}+\lambda_2k^{(2)})\cdot\phi)$. Thus, an
appropriate toy model for the double resonance is
\begin{equation}\label{eq:hammodeldble}
H_{doubleres} = H_0(J_{F1},J_{F2},J_F)
+\epsilon\sum_{l_1,l_2}g_{l_1,l_2}\cos(l_1\phi_{R1}+l_2\phi_{R2})~~.
\end{equation}
The coefficients $g_{l_1,l_2}$ are expressed in terms of the original
Fourier coefficients $h_k$. Now, contrary to the case of single
resonance, $H_{doubleres}$ has only one ignorable angle ($\phi_F$),
hence, besides the energy, only the action $J_F$ is integral of
motion.  Then, considering $J_F$ as a parameter, the dynamics of
$H_{doubleres}$ corresponds to a \textit{non-integrable} system with
two degrees of freedom. This is a general property of multiple
resonances, for which the Nekhoroshev normal form induces a
non-integrable dynamics. Availing no other restrictions than those
imposed by energy conservation, the dynamics near the junction can be
very chaotic (see, for example,~\cite{efthyhar-13},
\cite{gelsimo-13}). However, as discussed in \cite{bengall86} and
\cite{poshel-93}, energy conservation can still be used
in many cases to constrain the orbits consistently with the
Nekhoroshev theorem. As in the case of simple resonance, consider,
without loss of generality, the normal form dynamics induced by the
Hamiltonian Eq.~(\ref{eq:hammodeldble}) for (constant) $J_F=0$. The
normal form energy $E=H_{doubleres}$ is a constant of motion.  Thus,
the quantity $H_0(J_{R1},J_{R2},0)$ can only undergo ${\cal
  O}(\epsilon)$ oscillations around the value
$E=H_0(J_{R1},J_{R2},0)$. We then seek for conditions on $H_0$ such
that the manifold $E=H_0(J_{R1},J_{R2},0)$ be {\it bounded}, i.e. that
none of $J_{R1}$,$J_{R2}$ can take ${\cal O}(1)$ values while the
energy $E=H_0(J_{R1},J_{R2},0)$ still remains in the interval $E-{\cal
  O}\epsilon<H_0<E+{\cal O}(\epsilon)$. Subtracting an irrelevant
constant, consider values of the energy $E={\cal O}(\epsilon)$. We
have (for $J_F=0$):
\begin{eqnarray}\label{eq:quadform}
E&=&{1\over 2}\sum_{i=1}^3\sum_{j=1}^3 M_{ij}(I_*)
(k_i^{(1)}J_{R1}+k_i^{(2)}J_{R2})(k_i^{(1)}J_{R1}+k_i^{(2)}J_{R2})\nonumber\\
&=&\zeta_2=(J_{R1},J_{R2})Y(J_{R1},J_{R2})^T 
\end{eqnarray}
where $Y$ is the $2\times 2$ matrix
$$
Y = k^{(1,2)}M(I_*)(k^{(1,2)})^T
$$
with
\begin{displaymath}
k^{1,2} = 
\left(
\begin{array}{ccc}
k^{(1)}_1 &k^{(1)}_2 &k^{(1)}_3 \\
k^{(2)}_1 &k^{(2)}_2 &k^{(2)}_3
\end{array}
\right)
\end{displaymath}
The quadratic form (\ref{eq:quadform}) is positive definite when
$M(I_*)$ has three non-zero eigenvalues of equal sign, or two
eigenvalues of equal sign and one equal to zero.  In the first case,
the Hamiltonian $H_0$ will be called {\it convex}, and in the second
{\it quasi-convex}. In general, we give the following
definitions:\\ \\
{\it Convexity:} The n-degrees of freedom
Hamiltonian $H_0$ is convex at the point $I_*$ if there is a positive
constant $M$ such that for any $x\in R^n$, $x\neq 0$ we have
$|(M(I_*)x)\cdot x|\geq M$. \\ \\
{\it Quasi-convexity:} The
Hamiltonian $H_0$ is quasi-convex at the point $I_*$ if
$\omega(I_*)\neq 0$ and the only solution to the system
$\omega(I_*)\cdot x=0$ and $(M(I_*)x)\cdot x =0$ is $x=0$. \\ \\
We
leave to the reader as an exercise to demonstrate that when $H_0$ is
(quasi)convex at the point $I_*$, the $2\times 2$ matrix $Y$ of
Eq.(\ref{eq:quadform}) is positive definite (see also equation (171)
in~\cite{LaPlata}). Then, the equation $\zeta_2(J_{R1},J_{R2})=E$ is
the equation of an ellipse. For fixed value of $E={\cal O}(\epsilon)$,
both actions $J_{R1},J_{R2}$ are bounded by the fixed size (say, the
semi-major axis) of the ellipse.  The latter is of order
$\sqrt{\epsilon}$, hence the actions $J_{R1},J_{R2}$ are bounded in a
domain of size ${\cal O}(\sqrt\epsilon)$. On the contrary, at points
$I_*$ where (quasi-)convexity is not satisfied, the matrix $Y$ can be
positive-definite or not, depending on the particular resonant vectors
$k^{(1)},k^{(2)}$. Correspondingly, the equation
$\zeta_2(J_{R1},J_{R2})=E$ gives either an ellipse or a hyperbola. At
those junctions where we have hyperbolas, the actions $J_{R1},J_{R2}$
are unbounded along the asymptotes of the
hyperbolas\footnote{For example: $H_0=(I_1^2-I_2^2)/2+I_3$.  Then,
  $\omega_1=I_1,\omega_2=-I_2,\omega_3=1$, and $(M k)\cdot k =
  k_1^2-k_2^2$ which is not positive definite. Take the point
  $I_*=(1,1,0)$ corresponding to the double resonance
  $k^{(1)}=(1,1,0)$, $k^{(2)}=(1,0,-1)$. We obtain
  $J_1=J_{R1}+J_{R2}-J_F$, $J_2=J_{R1}+J_F$, $J_3=-J_{R2}-J_F$,
  implying $H_0=J_{R1}J_{R2}+J_{R_2}^2$. Then, the equation
  $E=J_{R1}J_{R2}+J_{R_2}^2={1\over 2}(J_{R1}+J_{R2})^2-J_{R1}^2$
  represents hyperbolas with the asymptotes $J_{R2}=0$ and
  $J_{R2}=-2J_{R1}$. Thus, even with energy $E=0$, the actions can
  move freely along the asymptotes without violating the constant
  energy condition.}.

In this case, however, a bound for the actions $J_{R1},J_{R2}$ via the
requirement $|H_0(J_{R1},J_{R2},0)|<{\cal O}(\epsilon)$ can still be
obtained using the cubic terms in the formula for $H_0$
(Eq.(\ref{eq:h0exp})). Without entering into details, we only mention
that such a bound exists when the Hamiltonian $H_0$ satisfies the
\textit{three-jet} condition:\\ \\ {\it Three-jet}: at the point $I_*$
we have $\omega(I_*)\neq 0$ and the only solution to the system of
equations
\begin{equation}\label{eq:threejet}
\omega(I_*)\cdot x=0,~~~(M(I_*)x)\cdot x =0,~~~
\sum_{i=1}^n\sum_{j=1}^n\sum_{l=1}^n
\left({\partial^3H_0\over\partial I_i\partial I_j\partial I_l}\right)_{I=I_*}x_ix_jx_l=0
\end{equation}
is $x=0$.  In the case $n=3$ the three-jet condition is generically
satisfied, as only coincidentally we can find a model $H_0$ in which all
three equations (\ref{eq:threejet}) be satisfied for some $x\neq
0$. However, when $n>3$ the fulfillment of the condition depends on
the choice of $H_0$ has to be checked case by case
(see~\cite{SchiGuz-13}).

Returning to the example of Figs.~\ref{fig:arnweb}(d), (e), we can easily
check the above conditions at the junctions A,B. We have
$A=(I_{*1},I_{*2},I_{*3})= (2\pi/3,2\pi/3,0)$,
$B=(I_{*1},I_{*2},I_{*3})=(2\pi/3,\pi/3,0)$. We saw already that The
Hessian matrix of $H_0$ is positive definite if $I_{*2}\geq
\pi/2$. Thus $H_0$ is convex in the case A. At B, instead, we have
$k_1=(3,1,0)$, $k_2=(3,0,-1)$, thus
$$
Y_B=\left(\begin{array}{cc}
     -2 &3\\
     3 &9
    \end{array}\right)
$$
with opposite sign eigenvalues $\lambda_{1,2}={1\over
  2}(7\pm\sqrt{157})$. This means that the quadratic form of
Eq.(\ref{eq:quadform}) yields hyperbolas (see figure \ref{fig:arndif}
below).

\subsection{Diffusion in the web of resonances}

\begin{figure}
\centering
\includegraphics[width=1.\textwidth]{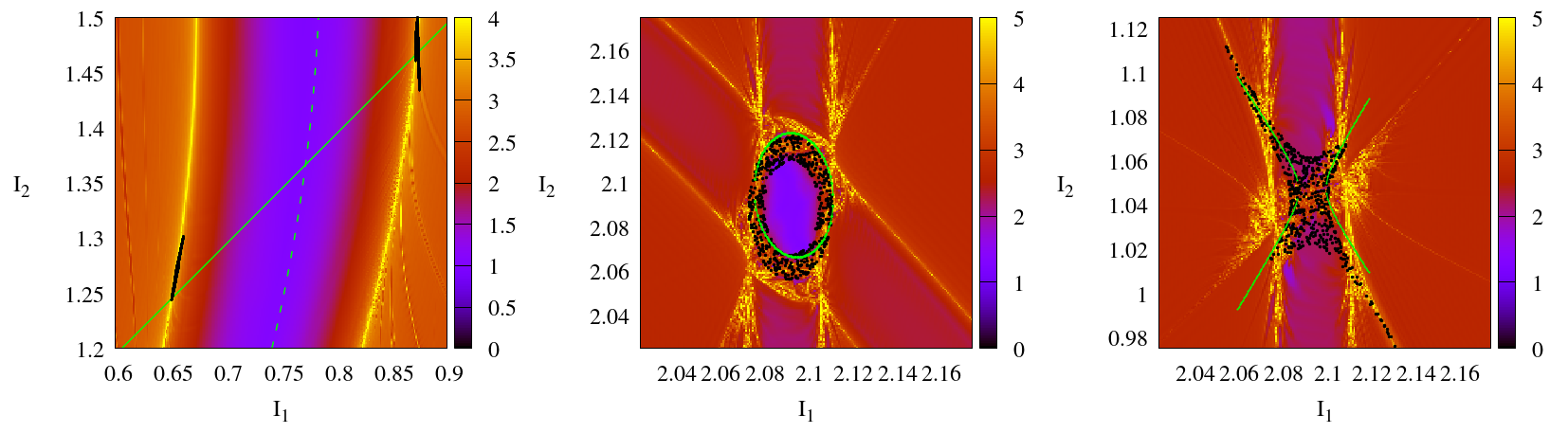}
\caption{(a) \textit{Left:} Arnold diffusion along a simple resonance
  in the model (\ref{eq:hamfr}) for $\epsilon=0.1$ (see
  text). \textit{Center:} Diffusion around the resonance junction A
  (quasi-convex domain). The ellipse represents the constant energy
  condition of Eq.~(\ref{eq:quadform}). \textit{Right:} Same as
  previously, but for the resonant junction B (non-convex, steep). The
  constant energy condition (\ref{eq:quadform}) now yields
  hyperbolas.}
\label{fig:arndif}
\end{figure}

We mentioned in section 2 that it is possible to prove the existence
of Arnold diffusion along the {\it simple} resonances of a priori
stable systems (see CG). The first numerical example of Arnold
diffusion in an a priori stable system similar to the one treated in
the examples above (but with $H_0=(I_1^2+I_2^2)+I_3$ satisfying
everywhere the quasi-convexity condition) was provided
by~\cite{legaetal-03}. Several more examples, including a spectacular
demonstration of the drift of the trajectories throughout the entire
Arnold web, were provided in~\cite{guzzoetal-05}.

Figure \ref{fig:arndif} (left) gives an example of the slow drift
along the resonance (1,1,0) in the model (\ref{eq:hamfr}) around the
point $I_*$ with $I_{1*}=0.77211...$, $I_{2*} = 1.3665$, $I_{3*}=0$,
for $\epsilon=0.1$. Using the FLI map, we first compute the borders of
the resonance (yellow). We then compute the plane of fast drift
crossing the chosen point $I_*$ (Eq.~(\ref{eq:plfd}), thin line in
Fig.~\ref{fig:arndif}).  Computing the FLI (for $t=1000$) for initial
conditions along this line, we obtain two points (on each separatrix
layer) where the FLI has a local maximum. The point of maximum on the
top right of the figure has co-ordinates $I_1= 0.87166$, $I_2 =
1.466054$. Taking trajectories in a very small square (of size
$10^{-5}$ in our case) around this point, and forward propagating
these trajectories, allows to observe their slow drift along the
separatrix layers of the resonance. The points in black in
Fig.~\ref{fig:arndif} correspond to only four such trajectories,
integrated up to a time $t=10^9$.  The trajectories are shown only
when returning to the same angular section $(\phi_1+\phi_2)\mod
2\pi=0$, and $\phi_3\mod 2\pi=0$ as the one for which the FLI was
computed (with a numerical tolerance $10^{-2}$). We notice that the
trajectories make an overall excursion in the action space of length
$\sim 0.5$ after this long integration time. Due to the selected
section, the trajectories yield points near the extrema of both
branches of the theoretical separatrix of the resonance (see previous
subsection), corresponding to the left and right groups of points in
Fig.~\ref{fig:arndif}, which are both produced by the {\it same}
trajectories. Besides the fast change in the resonant action $I_{R}$
(Eq.~(\ref{eq:restra})), we observe that the trajectories undergo a
slow change of the value of the adiabatic actions $I_{F1},I_{F2}$, a
fact making them to jump from one to a nearby plane of fast drift,
with all these planes parallel to the one shown in
Fig.~\ref{fig:arndif}. How to quantify these jumps will be discussed in
the next section.

The center and right panels of Fig.~\ref{fig:arndif} refer now to
chaotic trajectories around the resonant junctions A and B. We saw
that the quadratic form of the constant energy condition of
Eq.~(\ref{eq:quadform}) yields ellipses in the case of the point A,
while it yields hyperbolas in the case of the point B. Clearly, the
chaotic trajectories around the junction are governed by this
difference. In the case of the junction A, the normal form dynamics
impedes the chaotic trajectories to move beyond a layer of thickness
${\cal O}(e)$ around each ellipse. In the case of the junction B,
instead, the chaotic trajectories can have larger excursions by
following a path close to the asymptotes of the hyperbolas. In that
case, the trajectories are still limited around the resonant junction
due to the cubic terms in the Hamiltonian (\ref{eq:hamfr0}).

On the other hand, all predictions made by the normal form models are
valid up to an error determined by the exponentially small
\textit{remainder} of the normal form. More specifically, the
Nekhoroshev normal form has the form:
\begin{equation}\label{eq:nekhnf}
H_N = Z_N + R_N 
\end{equation}
where $Z_N$ is the normal form part and $R_N$ the remainder, with
$$
||R_N||={\cal O}\left(\exp((\epsilon_0/\epsilon)^b)\right)~~.
$$
Let $I_{Fi}$ (the `adiabatic actions') be the integrals of $Z_N$ (in the
3DOF case, $i=1,2$ in the case of a simple resonance, and $i=1$ in the
case of the double resonance). We have
\begin{equation}\label{eq:dotphir}
\dot{I}_{Fi}=-\partial R_N/\partial\phi_{Fi},~~~i=1,2 
\end{equation}
For the derivatives we have the estimate $||\dot{I}_{F_i}||={\cal O}
\left(\exp((\epsilon_0/\epsilon)^b)\right)$.  From this, we can
conclude that, although the actions $I_{Fi}$ cease to be integrals of
motion in the complete Hamiltonian, up to a given time $t$ the actions
$I_{F_i}$ can have excursions of length {\it bounded from above} by
$\Delta I_{F_i}< {\cal O}
\left(t\exp((\epsilon_0/\epsilon)^b)\right)$. This estimate yields
the local speed of Arnold diffusion, which can hence be measured using
the norm $||R_N||$. Another numerical test regards the comparison
between the numerically computed (by ensembles of trajectories) value
of the diffusion coefficient $D$, and the size of the remainder
$||R_N||$. Empirical fitting has given the law $D\sim||R_N||^3$ in the
case of simple resonances, and $D\sim||R_N||^2$ in the case of double
resonances.  Implementing the theory of Chirikov, instead, leads to
the estimate $D\propto ||R_N||^{2+\alpha}$, where the correction
$0<\alpha<1$ depends locally (in a simply-resonant domain) on the
detailed structure of the `layer resonances' determining the remainder
of the local Nekhoroshev normal form~\cite{cinetal-14}.

To unveil the detailed evolution of the variables $I_{F,i}(t)$ for any
trajectory one needs to solve the initial value problem for the
differential equations (\ref{eq:nekhnf}) up to any desired time $t$.
It turns out that, even availing the explicit expressions for a high
order truncation of the remainder $R_N$, in practice it is hard to try
to integrate the differential equations (\ref{eq:dotphir}) directly in
the computer. A good number of reasons impede us on this task,
starting from the fact that the remainder $R_N$ is actually a series,
whose representation in the computer is given by a truncated
trigonometric polynomial typically containing millions of terms. This
is an expression hard to deal with not only numerically, but also in
any theoretical attempt to establish the existence of phase space
objects (e.g. manifolds like the ones of Fig.~\ref{fig:arnsche}) having
the role of drivers of Arnold diffusion.
\footnote{While drifting along a simple resonance, a chaotic
  trajectory will eventually reach a multiple resonance domain. For
  some time, the trajectory then behaves as shown in the middle and
  right panels of Fig.(\ref{fig:arndif}).  To demonstrate Arnold
  diffusion requires, however, showing that the trajectory will
  eventually exit from the multiple resonance, continuing to drift
  along the same exit simple resonance as the entry one, or choosing a
  different exit resonance. The lack of proof, in a priori stable
  systems, of the existence of a mechanism guaranteeing that these
  transitions will take place, is known as the `large gap
  problem'~\cite{delshetal-06},\cite{delshetal-16}. The existence of
  orbits undergoing long excursions in a priori stable systems, but
  far from double resonances, is demonstrated in \cite{kaloetal-16}.}

On the other hand, we can always attempt to {\it model} the dynamics
of itself the remainder $R_N$. As discussed in the sequel, such a
modeling is possible and leads to a way more tractable expression
$R_N^{(model)}$. Using $R_N^{(model)}$ we can then probe and visualize
most phenomena related to Arnold diffusion. In particular, we can
unravel the `jumps' in action space (similar as in
Fig.~\ref{fig:arnoldex}(d)) undergone by the weakly chaotic trajectories
within the layers of a selected resonance.  We can also predict and
model the size of these jumps. Finally, we can identify the fastest
drifting trajectories and monitor how close their speed is to the
theoretical upper bound provided by the Nekhoroshev theorem $\Delta
I_{F_i}(max)= t\sup_{{\cal D}_*}|\partial R_N/\partial\phi_{Fi}|$ (see
examples in the next section).

\section{Construction of the Nekhoroshev normal form: semi-analytical estimates}

\subsection{Construction of the Nekhoroshev normal form}
It was mentioned before that most semi-analytical results on the
quantification of the Arnold diffusion follow after the appropriate
construction of a local Nekhoroshev normal form in a selected domain
${\cal D}_*$ around some point $I_*\in\mathbb{R}^n$ of the action
space of the problem. We here summarize the method implemented in
\cite{efthy-08},\cite{efthyhar-13},\cite{cinetal-14},\cite{GEP}, for 
an efficient computation of the Nekhoroshev normal form. We assume 
a n-DOF system with Hamiltonian
\begin{equation}\label{eq:themodel}
H(I,\phi) = H_0(I) + \epsilon H_1(I,\phi)~~~,
\end{equation}
satisfying the properties enumerated below. 

\subsubsection{Analyticity}
We assume that there is an open domain ${\cal I}\subset\mathbb{R}^3$
and real constants $\rho>0,\sigma>0$ such that for all points $I_*\in{\cal I}$ and 
all complex quantities $J_i\in\mathbb{C}$, $i=1,\ldots n$ satisfying $|J_i|<\rho$ 
the following properties hold true:\\
\\
i) the function $H_0$ can be expanded as a convergent Taylor series
\begin{equation}\label{h0exp}
H_0=H_{0*}+\omega_*\cdot J
+ {1\over 2}\sum_{i=1}^n\sum_{j=1}^n M_{ij*}J_iJ_j +\ldots
\end{equation}
where $\omega_*=\nabla_I H_0(I_*)$ and $M_{ij*}$ are the elements of the Hessian 
matrix of $H_0$ at $I_*$, denoted by $M_*$.

ii) For all $I_*\in {\cal I}$, $H_1$ admits a Fourier expansion
\begin{equation}\label{h1four}
H_1=\sum_k h_k(I_*+J)\exp(ik\cdot\phi)
\end{equation}
analytic in the domain 
\begin{equation}\label{eq:caldstar}
{\cal D}(I_*)=\left\{I_i=I_{i*}+J_i,|J_i|<\rho,\Re(\phi_i)\in\mathbb{T}, 
|\Im(\phi_i)|<\sigma, i=1,\ldots n\right\}~~.
\end{equation}
The analyticity of the function $H_1$ in the domain ${\cal D}$ implies that all the 
coefficients $h_k$ can be expanded in convergent Taylor series around $I_*$ as 
\begin{equation}\label{eq:hkexp}
h_k=h_{k*}+\nabla_{I_*}h_k\cdot J
+ {1\over 2}\sum_{i=1}^n\sum_{j=1}^nh_{k,ij*}J_iJ_j +\ldots
\end{equation}

\subsubsection{Book-keeping}
Due to the analyticity of $H_1$, the Fourier coefficients $h_k$ in the domain 
${\cal D}(I_*)$ decay exponentially, that is, there are positive constants $A$, 
$\sigma$ such that 
\begin{equation}\label{eq:expdecay}
\sup_{\cal D_*}|h_k(I)|<A e^{-|k|\sigma}
\end{equation}
Taking the exponential decay into account, we then split the Fourier harmonics in 
groups with the wave number satisfying $(s-1)\leq|k|<sK-1$, $s=1,2,\ldots$, and  
 \begin{equation}\label{eq:kappa}
K = -{1\over\sigma}\log(\rho_0)~~,
\end{equation} 
where $\rho_0$ is the size of the domain around the point $I_*$ where the normal 
form is to be computed, i.e., $|J_i|<\rho_0$. For resonant constructions of any 
multiplicity it is convenient to take $\rho_0 = {\cal O}(\sqrt{\epsilon})$. 
Introducing a `book-keeping' symbol $\lambda$, with numerical value $\lambda=1$, 
the Hamiltonian can then be split in ascending powers of $\lambda$: 
\begin{eqnarray}\label{hamexpf}
H=H^{(0)}(J,\phi)&=&Z_0+\sum_{s=1}^{\infty}\lambda^s
H^{(0)}_s(J,\phi;\epsilon)
\end{eqnarray}
where
$$
Z_0=\omega_*\cdot J
$$
and 
\begin{equation}\label{h0s}
H^{(0)}_s = \sum_{\mu=1}^s
\sum_{k=K'(s-\mu)}^{K'(s-\mu+1)-1} H^{(0)}_{\mu,k}(J)\exp(ik\cdot\phi)
\end{equation}
where $H^{(0)}_{\mu,k}(J)$ are polynomials containing terms of degree
$\mu-1$ or $\mu$ in the action variables $J$. In the $n=3$ cases dealt with in the 
numerical examples of this article, we have, in particular:
$$
H^{(0)}_{\mu,k}(J)=\epsilon
\sum_{\mu_1=0}^{\mu-1}~~
\sum_{\mu_2=0}^{\mu-1-\mu_1}~~
\sum_{\mu_3=0}^{\mu-1-\mu_1-\mu_2}
{1\over\mu_1!\mu_2!\mu_3!}
{\partial^{\mu-1}h_{1,k}(I_*)\over
\partial^{\mu_1}I_1\partial^{\mu_2}I_2\partial^{\mu_3}I_3}
J_1^{\mu_1}J_2^{\mu_2}J_3^{\mu_3}
$$
if $|k|>0$, or
$$
H^{(0)}_{\mu,k}(J)=
\sum_{\mu_1=0}^{\mu}~~
\sum_{\mu_2=0}^{\mu-\mu_1}~~
\sum_{\mu_3=0}^{\mu-\mu_1-\mu_2}
{1\over\mu_1!\mu_2!\mu_3!}
{\partial^{\mu}H_0(I_*)\over
\partial^{\mu_1}I_1\partial^{\mu_2}I_2\partial^{\mu_3}I_3}
J_1^{\mu_1}J_2^{\mu_2}J_3^{\mu_3}
$$
$$
~~~~~~~~+\epsilon\sum_{\mu_1=0}^{\mu-1}~~
\sum_{\mu_2=0}^{\mu-1-\mu_1}~~
\sum_{\mu_3=0}^{\mu-1-\mu_1-\mu_2}
{1\over\mu_1!\mu_2!\mu_3!}
{\partial^{\mu-1}h_{1,0}(I_*)\over
\partial^{\mu_1}I_1\partial^{\mu_2}I_2\partial^{\mu_3}I_3}
J_1^{\mu_1}J_2^{\mu_2}J_3^{\mu_3}
$$
if $k=0$.  In all the above expressions, the superscript $(0)$
means `the starting Hamiltonian of the iterative normalization
process'. This is simply the original Hamiltonian re-organized in
powers of the book-keeping symbol $\lambda$. Subscripts (as e.g. $s$
in the functions $H^{(0)}_s(J,\phi;\epsilon)$) mean terms book-kept
with the power $\lambda^s$. In physical terms, this can be interpreted
as `terms of the s-th order of smallness'. All expressions in the
initial and in subsequent normalization steps are finite, i.e., they
are trigonometric polynomials easily represented in the computer's
memory via an indexing function. The maximum `book-keeping' order
$N_{tr}$ adopted in the normalization algorithm is called the
truncation order.

\subsubsection{Resonant module}
Following the definitions given in subsection 4.2, the point $I_*$, 
and its corresponding frequency vector $\omega_*=\omega(I_*)$, are called 
`$M-$tuple resonant' (with $0\leq M\leq n-1$) if there can be found $M$ 
linearly independent non-zero integer vectors $k^{(i)}$, $i=1,\ldots M$ 
such that $k^{(i)}\cdot\omega_* = k^{(i)}\cdot\omega(I_*)=0$ for all 
$i=1,\ldots,M$. When a point $I_*$ is $M-$tuple resonant, there are many 
harmonics $\cos(k\cdot\phi)$ with $|k|\neq 0$ in the Hamiltonian which 
cannot be normalized since their elimination would involve a divisor exactly 
equal to zero. The set of all possible wavevectors $k$ such that 
$k\cdot\omega_*=0$ is called the \textit{resonant module} at the point 
$I_*$. Since checking numerically the condition $k\cdot\omega_*=0$,  
with $\omega_*\in\mathbb{R}^n$, is sensitive to round-off errors, a 
convenient way to define the resonant module, which involves only operations 
among integer numbers, is by use of the concept of `pseudo-frequency' vector. 
This is defined as follows: if $\omega_*$ is $M-$tuple resonant with $M\geq 1$, 
choose $M$ non-zero linearly independent integer vectors $k^{(i)}$, $i=1,
\ldots,M$ such that $k^{(i)}\cdot\omega(I_*)=0$. Then, there exist 
$n-M$ non-zero integer vectors $m^{(j)}$, $j=1,\ldots,n-M$ such
that $k^{(i)}\cdot m^{(j)}=0$ for all possible pairs $i,j$. To define
these vectors, solve the $n-M$ systems of linear equations given by
\begin{eqnarray}\label{eq:pseudo}
&~&k^{(1)}_1q_1^{(j)}+k^{(1)}_2q_2^{(j)}+\ldots+k^{(1)}_Mq_M^{(j)}
=-k^{(1)}_{M+j} \nonumber\\
&~&k^{(2)}_1q_1^{(j)}+k^{(2)}_2q_2^{(j)}+\ldots+k^{(2)}_Mq_M^{(j)}
=-k^{(2)}_{M+j} \\
&~&\ldots\nonumber\\
&~&k^{(M)}_1q_1^{(j)}+k^{(M)}_2q_2^{(j)}+\ldots+k^{(M)}_Mq_M^{(j)}
=-k^{(M)}_{M+j}\nonumber 
\end{eqnarray}
for $j=1,\ldots,n-M$. The solutions give vectors
$q^{(j)}=(q_1^{(j)},\ldots,q_M^{(j)},\delta_{M+1,M+j},\ldots,\delta_{n,M+j})$
with rational components. Multiplying the vector $q^{(j)}$ with the
maximal common divisor of all its components yields the j-th
pseudo-frequency vector $m^{(j)}$.

We can now determine which harmonics $\cos(k\cdot\phi)$ to be excluded
from the normalization process. The set of all integer vectors $k$
corresponding to the excluded harmonics is called the {\it resonant
  module} ${\cal M}(k^{(1)},\ldots,k^{(M)})$ defined as:
\begin{equation}\label{eq:resmod}
{\cal M}(k^{(1)},\ldots,k^{(M)})=\left\{
\begin{array}{ll}
\{k=(0,0,\ldots,0)\} 
&~~\mbox{if}~M=0\\
\{k\in\mathbb{Z}^n: k\cdot m^{(j)}=0~~~\mbox{for all}~j=1,\ldots,n-M\} 
&~~\mbox{if}~M>0
\end{array}\right.
\end{equation}
where $m^{(j)}$, $j=1,\ldots,n-M$ are the pseudo-frequency vectors determined through 
Eq.(\ref{eq:pseudo}).

Note that, even when the origin of the expansion $I_*$ is non-resonant, 
i.e., when $M=0$, arbitrarily close to it there can be found $M-$tuple resonant 
points of any multiplicity $M>0$. This is a consequence of the fact that 
resonances are dense in the action space (see the examples in \cite{LaPlata}). 
Whenever the non-resonant vector $\omega_*$ is `close' to a low-order M-tuple 
resonant vector $\Omega$, in the sense that $|\omega_*-\Omega|<\alpha$ 
with $\alpha$ small, and the wavevectors $k$ satisfying $k\cdot\Omega$ 
are of order $|k|$ smaller than the `cut-off' order (see below), we say 
to be in a `near-resonance' case. In this case too, we may wish to avoid 
the presence in the series of those divisors $k\cdot\omega_*$ for which 
$k\cdot\Omega=0$. We then define the resonant module as above, but using 
$\Omega$ in the place of $\omega_*$.

\subsubsection{Hamiltonian normalization}
We consider a sequence of normalizing canonical transformations 
$$
(\phi,J)\equiv(\phi^{(0)},J^{(0)})\rightarrow(\phi^{(1)},J^{(1)})
\rightarrow(\phi^{(2)},J^{(2)})\rightarrow\ldots
$$
leading to re-express the Hamiltonian, after $r$ normalization steps, 
in new canonical variables $(\phi^{(r)},J^{(r)})$ such that
\begin{equation}\label{eq:nfcompo}
H(\phi((\phi^{(r)},J^{(r)}))=
Z^{(r)}((\phi^{(r)},J^{(r)});\lambda,\epsilon)
+R^{(r)}(\phi^{(r)},J^{(r)};\lambda,\epsilon) ~~.
\end{equation}
The functions $Z^{(r)}(J^{(r)},\phi^{(r)};\lambda,\epsilon)$ and
$R^{(r)}(J^{(r)},\phi^{(r)};\lambda,\epsilon)$ are called the normal
form and the remainder respectively. The normal form is a finite
expression which contains terms up to order $r$ in the book-keeping
parameter $\lambda$. By definition, these are terms belonging to the
resonant module ${\cal M}(k^{(1)},\ldots,k^{(M)})$. The remainder,
instead, is a convergent series containing terms of order
$\lambda^{r+1}$, including all possible harmonics.

To compute the normalizing transformation, we use the composition of
Lie series with generating functions $\chi_1,\ldots,\chi_r$. Denote
$Q=(\phi,J)\equiv Q^{(0)}$. The normalizing transformation is:
\begin{equation}\label{lietra}
Q^{(r)}=\exp(-L_{\chi_1})\exp(-L_{\chi_2})\ldots\exp(-L_{\chi_r})Q \\
\end{equation}
The generating functions are determined recursively, by solving, for\\
$n_r=0,\ldots,r-1$ the homological equations:
\begin{equation}\label{homo}
\{\omega_*\cdot J^{(n_r+1)},\chi_{n_r+1}\}+\lambda^{n_r+1}
\tilde{H}^{(n_r)}_{n_r+1}(J^{(n_r+1)},\phi^{(n_r+1)})=0
\end{equation}
where 
\begin{equation}\label{hr}
H^{(n_r)} = \exp(L_{\chi_{n_r}})H^{(n_r-1)}~~.
\end{equation}

\subsubsection{Optimal remainder}
Basic normal form theory (see~\cite{LaPlata}) establishes that
the above normalization process has an \textit{asymptotic} character.
Namely, i) the domain of convergence of the remainder series $R^{(r)}$
shrinks as the normalization order $r$ increases, and ii) the size
$||R^{(r)}||$ of $R^{(r)}$, where $||\cdot||$ is a properly defined
norm in the space of trigonometric polynomials, initially decreases,
as $r$ increases, up to an optimal order $r_{opt}$ beyond which
$||R^{(r)}||$ increases with $r$.  In the {\it Nekhoroshev regime},
one has $||Z^{(r_{opt})}||>>||R^{(r_{opt})}||$. Hence, the normal form
obtained at the order $r_{opt}$ best unravels the dynamics, which is
given essentially by the Hamiltonian flow of $Z^{(r_{opt})}$ slightly
perturbed by $R^{(r_{opt})}$. Furthermore, the optimal normalization
order $r_{opt}$ depends on $\epsilon$ via an inverse power-law
(\cite{efthy-08}\cite{efthyhar-13}), namely
\begin{equation}\label{ropt}
r_{opt}\sim\epsilon^{-a}~~,
\end{equation}
for some positive exponent $a$ depending on the multiplicity of the
resonance around which the normal form is computed. The leading terms
in the optimal remainder function are $O(\lambda^{r_{opt}+1})$. Due to
the book-keeping relation (\ref{eq:kappa}), the terms of order
$\lambda^{r_{opt}}$ have size estimated as $e^{-\sigma K_{opt}}$,
where
\begin{equation}\label{kopt}
K_{opt}(\epsilon)=K'r_{opt}(\epsilon)
\end{equation}
is called the \textit{Nekhoroshev cut-off} order. Then, $K_{opt}\sim
K'\epsilon^{-a}$, implying:
\begin{equation}\label{ropt}
||R^{(r_{opt})}||\sim
\epsilon^{1/2}\exp\left({-K'\sigma\over\epsilon^{a}}\right)
\end{equation}
i.e., the remainder at the optimal normalization order is
exponentially small in $1/\epsilon$.

\begin{figure}[!ht]
\centering
\includegraphics[width=0.6\textwidth]{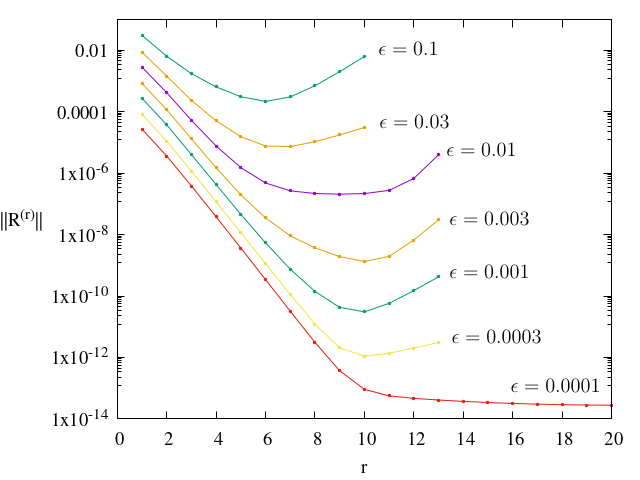}
\caption{Size of the remainder as a function of the normalization order $r$ 
for various values of $\epsilon$. The value of $r$ at the minimum of each 
curve corresponds to the optimal normalization order. Note that the optimal 
order is higher than 20 in the case $\epsilon=0.0001$.}
\label{fig:optrem}
\end{figure}

In practice, to specify the optimal normalization order, after
performing all the above symbolic computations with the aid of a
computer program, we proceed as follows: we set the truncation order
$N_t$ to be several orders larger than the maximum reached
normalization order $r$. Then, we compute the truncated-norm estimates
\begin{equation}\label{remnorm}
||R^{(r)}||_{W^{(r)}} = \sum_{s=r+1}^{N_t}\sum_m\sup|R^{(r)}_s|_{W^{(r)}}
\end{equation}
where $\sup|R^{(r)}_s|$ means the sup norm of the s-th book-keeping
term of the truncated remainder over a domain of interest $Q^{(r)}\in
W^{(r)}$ where the r-th step canonical variables. To this end, we
first probe numerically that $W^{(r)}$ is smaller than the convergence
domain for the r-th step normalization.  We then verify the asymptotic
character of the sequence $||R^{(r)}||_{W^{(r)}}$, for
$r=1,2,3,\ldots$. That is, for $\epsilon$ sufficiently small,
initially (at low orders) $||R^{(r)}||_{W^{(r)}}$ decreases as $r$
increases, up to the optimal order $r_{opt}$ at which
$||R^{(r_{opt})}||_{W^{(r_{opt})}}$ reaches a minimum. Then, for
$r>r_{opt}$, $||R^{(r)}||_{W^{(r)}}$ increases with $r$. This behavior
is exemplified in Fig.\ref{fig:optrem}, referring to the normal form
computed for the data of the simple resonance corresponding to the
left panel of Fig.~\ref{fig:arndif}.

\subsection{Removal of deformation effects}
We have seen that, at the optimal order, the adiabatic actions
$I^{(r_{opt})}_{Fi}$ are integrals of the normal form dynamics,
while in the full Hamiltonian they undergo exponentially small time
variations due to the exponentially small optimal remainder.  One
important effect, which impedes to measure the real speed of the
variations of the adiabatic action variables is
\textit{deformation}. Consider the inverse of the transformation
(\ref{lietra}) at optimal order:
\begin{equation}\label{lietrainv}
Q=\exp(L_{\chi_r})\exp(L_{\chi_{r-1}})\ldots\exp(L_{\chi_1})Q^{(r_{opt})} \\
\end{equation}
Due to the relation $\exp(L_{\chi_1})Q^{(r_{opt})}=Q^{(r_{opt})}+\{Q^{(r_{opt})},\chi_1\}
+\ldots$, as well as the fact that $\chi_s={\cal O}(\rho_0^s)$, we have that 
\begin{equation}\label{qqr}
Q =  Q^{(r_{opt})} + {\cal O}(\rho_0)
\end{equation}
Furthermore, for resonant normal forms, we saw that $\rho_0={\cal
  O}(\epsilon^{1/2})$.  Thus, we find that even while the adiabatic
actions $I^{(r_{opt})}_{Fi}$ undergo a very slow time variation
(including drift), in the original variables this variation is
completely hidden in a ${\cal O}(\epsilon^{1/2})$ oscillation, due
entirely to the canonical transformation linking old with new
variables. Since, without knowledge of the normalizing transformation,
we are forced to deduce all the information on the behavior of the
system by numerical experiments performed using the original
variables, this implies that we have to recover the drift by removing
all the noise induced by these large amplitude, but irrelevant for
dynamics, oscillations.

\begin{figure}[!ht]
\centering
\includegraphics[width=1.\textwidth]{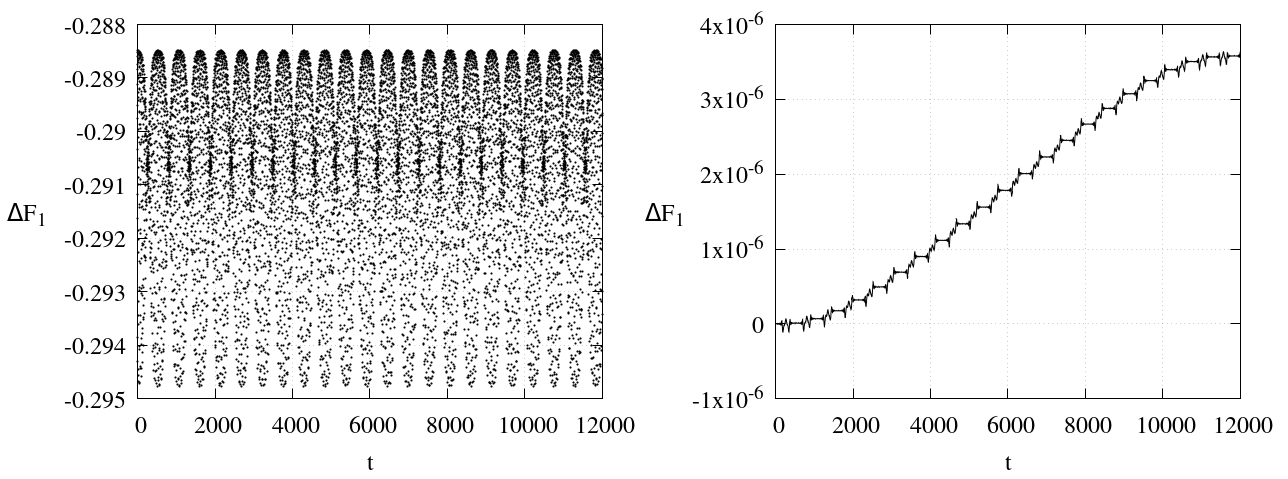}
\caption{Evolution of the adiabatic action $J_F$ along a simple resonance in the 
model (\ref{eq:hamfr}). \textit{Left}: numerical trajectory. \textit{Right:}, the 
same trajectory, but plotted in the optimal canonical variable $J_F^{(r_{opt})}$ 
(see text).}
\label{fig:drift}
\end{figure}
Being able to compute the optimal normalizing transformation, allows,
instead to spectacularly remove the deformation effect and easily obtain 
(and measure) the underlying drift of the adiabatic action
variables. Figure \ref{fig:drift} shows the removal of the deformation
in the case of a trajectory undergoing Arnold diffusion in the model
(\ref{eq:hamfr}) and with initial condition as in
Fig.~\ref{fig:arndif}. Recall that to visualize the drift using the
original variables in that case has required an extremely long
integration time $t=10^9$. For quite shorter times, instead, ($t=10^4$
in Fig.~\ref{fig:drift}) the drift of the unique adiabatic action of
the problem, measured by $\Delta {F,1} = I_F(t)-I_F(0)$ is completely
hidden in a oscillation of size $~0.2$ (Fig.~\ref{fig:drift}, left),
and thus impossible to measure with numerical experiments up to the
time $t=10^4$. If, instead, we pass all the numerical data $Q(t)$ of
the trajectory through the optimal normalizing transformation
(Eq.~(\ref{lietra})), we obtain the evolution of the optimal variable
$\Delta {F,1}^{(r_{opt})} = I_F^{(r_{opt})}(t)-I_F^{(r_{opt})}(0)$,
shown in Fig.~\ref{fig:drift}, right. Now, the drift is clearly
demonstrated, and its local velocity can be measured by a simple 
fitting to the data. In fact, as discussed in the next subsection, the 
drift in the action space is not necessarily monotone, and 
$\Delta {F,1}^{(r_{opt})}(t)$ may exhibit both an increase or decrease 
at different intervals of time. At any rate, the ability to remove the 
deformation effect can be exploited in the modeling of the evolution of 
the adiabatic action variables, as discussed in the next subsection.

\subsection{Modeling the jumps in the adiabatic action variables}
We have mentioned that it is possible to prove the occurrence of the Arnold mechanism 
in a priori stable systems only in the case of simple resonances (CG). We will now 
discuss how to model the evolution of the adiabatic action variables, including the 
jumps similar in nature as those of the original Arnold model, using, however, the 
information encapsulated in the remainder at the optimal normalization order. 
Consider an optimal Hamiltonian of the form (\ref{eq:nfcompo}) obtained by 
normalization around a simply-resonant point $I_*$. 

Following~\cite{GEP}, to simplify all notations, denote as $H^N$ 
(the `Nekhoroshev normal form') the Hamiltonian $H^{(r_{opt})}$, 
depending on the resonant action-angle variables $(S,\sigma)\equiv \left(J_R^{(r_{opt})},
\phi_R^{(r_{opt})}\right)$ and the $n-1$ adiabatic action variables conjugate to fast 
angles $(F,\phi)\equiv\left(J_F^{(r_{opt})},\phi_F^{(r_{opt})}\right)$ 
(see section 3). With the new notation, we have
\begin{equation}
H^N= h(F,S)+\epsilon f^N (F,S,\sigma)+ r^N (F,S,\sigma,\phi)\label{HN}~~~.
\end{equation}
The (simply-resonant) normal form is
\begin{equation}\label{HNappr} 
\overline H^N= h(F,S)+\epsilon f^N(F,S,\sigma)~.
\end{equation} 
The remainder $r^N$ is provided as a Taylor-Fourier series: 
\begin{equation}
r^N = \sum_{m\geq 0}\sum_{\nu \in {\Bbb Z}^d}
\sum_{k\in {\Bbb Z}^{n-d}}r^m_{\nu,k}(F)(S-S_*)^m 
e^{i\nu \cdot \sigma+ik\cdot \phi}
\label{remainder}
\end{equation}
expanded at a suitable $S_*$, with computer-evaluated truncations
involving a large number (typically $10^7$ to $10^8$) terms.

To define the resonant normal form dynamics, as in section 3 we first
expand $\overline H^N$ at the values of the actions $(F_*,S_*)$
identifying the center of the resonance, where
\begin{equation}\label{centerFS}
{\partial h\over \partial S}(S_*,F_*)=0 ,
\end{equation}
Then
\begin{equation}\label{HB0}
\overline H= \overline H_0+...\ \ ,\ \ \overline H_0=\omega_*\cdot \hat F
+ {A\over 2}\hat S^2 + \hat S  B\cdot \hat F +
{1\over 2} C\hat F\cdot \hat F + \epsilon v(\sigma)
\end{equation}
where $\hat F=F-F_*$, $\hat S=S-S_*$, $A\in {\Bbb R}$, $\omega_*,B\in
{\Bbb R}^{n-1}$, $C$ is a $(n-1)\times(n-1)$ square matrix and
$v(\sigma)$ is a trigonometric function depending parametrically on
$S_*(I_*),F_*(I_*)$.  The actions $\hat F$ are the constants of motion
for the Hamiltonian flow of $\overline H_0$.

Consider, now, the family of curves $\hat{S}(u;\alpha)$, for different
$a$, given by
\begin{equation}
\hat S= \sqrt{\epsilon}s_\alpha(\sigma) =  
\pm \sqrt{\epsilon}\sqrt{{2\over \norm{A}}(M(1+\alpha)-v(\sigma))}
\label{hats}
\end{equation}
where $M=\max_{\sigma\in [0,2\pi]} v(\sigma)$, and $\alpha$ is the energy of 
the pendulum Hamiltonian (equal to $\overline H_0$ for $\hat F=0$):
\begin{equation}\label{eq:alphapend}
a={A\over 2}\hat S^2+\epsilon v(\sigma)=\epsilon M(1+\alpha)~~.
\end{equation}
Since $\overline H_0$ has the structure of a pendulum Hamiltonian, we
can attempt to implement the Melnikov approximation, introduced in
section 2, in order to compute the jumps in the variables $F$ over one
complete homoclinic transition of the variables $(\hat{S},\sigma)$,
assigning to the remainder $r^N$ (Eq.~(\ref{remainder})) the role of
the coupling term between the resonant variables $(S,\sigma)$ and the
remaining variables $(F,\phi)$. Since $\dot{F_j}=-\partial
r^N/\partial\phi_j$, the Melnikov approximation will then consist of
estimating the variation $\Delta F_j(T)=F_j(t)-F_j(0)$ after a time
$T$ via the integral
\begin{equation}\label{eq:dftmel}
\Delta F_j(T) =-\sum_{m,\nu,k} \int_0^{T}  i k_j r^m_{\nu,k}(F(t))\hat S(t)^m 
e^{i\nu \sigma(t) + ik\cdot \phi(t)}dt
:= \sum_{m,\nu,k} \Delta F_{j,T}^{m,\nu,k}~~.
\end{equation}
where the true solution $(F(t),S(t),\sigma(t),\phi(t))$ in the r.h.s
of the integrals (\ref{eq:dftmel}) will be substituted by the
approximate solution under the flow of the normal form $\overline H_0$
$$
(F_*,S^0(t),\sigma^0(t),\phi^0(t))=(F_*,S_*,0,0)+
(0, \hat S^0(t),\sigma^0(t),\phi^0(t))
$$
where $(0,\hat S^0(t),\sigma^0(t),\phi^0(t))$ is a solution of Hamilton's 
equations of $\overline H_0$. 

Contrary to the simple model of section 2, it is important to recall
that the number of Melnikov integrals to compute in (\ref{eq:dftmel})
are of the same order as the number of remainder terms ($10^7$ to
$10^8$), thus the computation is hardly tractable in practice.
However, we get an enormous simplification of the problem noticing
that, out of all these integrals, only few ($\sim 10^3$) really contribute
to the result. To this end, we first observe that representing $\hat
S^0(t)$ parametrically as a function of $\sigma^0(t)$, for fixed
$\alpha$, allows to change the integration variable in
(\ref{eq:dftmel}) from $t$ to $\sigma$:
\begin{equation}\label{eq:melnikint}
\Delta F_{j,T}^{m,\nu,k}(T)\simeq 
\Delta^0 F_{j,T}^{m,\nu,k}(T) \hskip -0.2 cm = 
- i k_j {{r^m_{\nu,k}(F_*) \epsilon^{m-1\over 2}\over A}e^{ik\cdot\phi(0)}}
\hskip -0.2 cm
\int_0^{\sigma^0(T)}\hskip -0.6cm  [s_\alpha(\sigma)]^{m-1} 
e^{i \theta(\sigma)}d\sigma
\end{equation}
where the phase $\theta(\sigma)$ is defined by:
$$
\theta(\sigma)= {{\cal N}\sigma +{\Omega\over A\sqrt{\epsilon}}
\int_0^{\sigma} {dx\over s_\alpha(x)} } 
$$
with 
\begin{equation}
{\cal N}= \nu +k\cdot B/A,~~~~ \Omega =k\cdot \omega_*~~~. 
\label{calnomega} 
\end{equation}
Then, invoking the principle of stationary phase, it is clear that only 
integrals involving a slow variation of the phase $\theta(\sigma)$ over 
a time $T_\alpha$, representing the period of one homoclinic transition, 
will be important in the computation of the jumps via the
Eq.~(\ref{eq:melnikint}).

To make this argument more explicit, assume that the lowermost order
terms in the resonant normal form (for $\hat F=0$) have the form of
the pendulum Hamiltonian:
\begin{equation}
H_{pend} = {|A|\over 2}{\hat S}^2 + \epsilon\beta\cos\sigma +...
\label{hpend}
\end{equation}
where, for simplicity, we set $\epsilon,\beta>0$.  Consider a
remainder term labeled by the integers $(m,\nu,k)$ in
Eq.~\eqref{eq:melnikint}. Using the approximation (\ref{hpend}), and
setting $\alpha = 0$ (separatrix solution), the function
$\theta(\sigma)$ for the term in question can be approximated by:
\begin{equation}
\theta(\sigma)\approx \theta_0+{\cal N}\sigma +{\cal W}\ln\tan(\sigma/4),~~~
{\cal W} = {\Omega\over\sqrt{|A|\beta\epsilon}}~~.
\label{thsigpend}
\end{equation}
where ${\cal N}$ and $\Omega$ are given by Eq.~\eqref{calnomega},
hence, they depend only on the term labels $\nu,k$. From
Eq.~\eqref{thsigpend}, we obtain
\begin{equation}\label{thetaprim}
\theta'(\sigma) \approx {\cal N} + \frac{{\cal W}}{2} \frac{1}{\sin
  (\sigma/2)}
\end{equation}
Therefore, one has $\lim_{\sigma\rightarrow 0}W\theta'(\sigma) =
\lim_{\sigma\rightarrow 2\pi}W\theta'(\sigma) = +\infty$, and since
$\theta'(\sigma)$ is a function symmetric with respect to $\pi$ and
monotonically decreasing (increasing) in $[0,\pi)$ ($(\pi,2\pi]$),
there exists a minimum of the function at $\sigma= \pi$ of value
$\theta'(\sigma) = {\cal N} + {\cal W}/2$.  Thus, $\theta'(\sigma)$
has zeroes (stationary points) $\sigma_c = \pi\pm\Delta\sigma_c$, with
$0<\Delta\sigma_c<\pi$, if and only if the minimum value
$\theta'(\pi)$ is negative. This lead to the following condition:
\begin{equation}\label{eq:stati}
  \mathrm{The\,\,term\,\,defined\,\,by\,\,}(m,\nu,k)
  \mathrm{\,\,is\,\,stationary}
  \iff
{\cal N}\cdot{\cal W}<0\,\,\mathrm{and}\,\,|{\cal N}|>{|{\cal W}|\over 2}
\end{equation}

\begin{figure}
\centering
\includegraphics[width=1.0\textwidth]{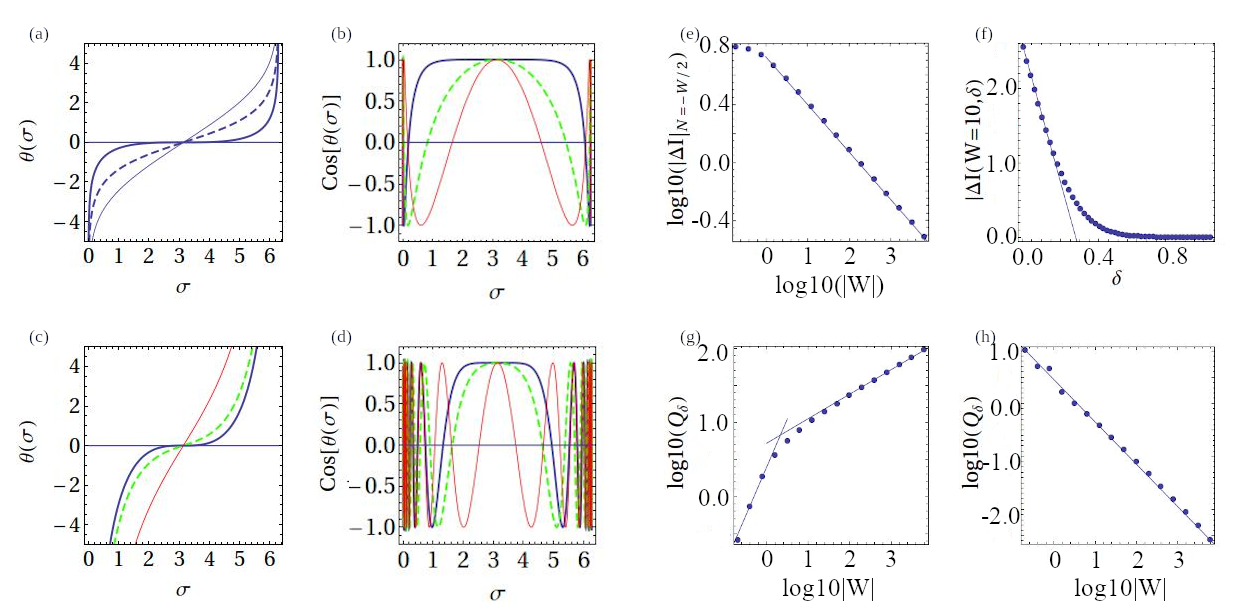}
\caption{(a) The function $\theta(\sigma)$ (equation~\ref{thsigpend})
  written as $\theta(\sigma)={\cal
    W}[0.5(1-\delta)](\sigma-\pi)+\ln\tan(\sigma/4))$ for ${\cal W}=1$
  and $\delta=0$ (thick blue), $\delta=1$ (dashed green), or
  $\delta=2$ (thin red). The corresponding curves
  $\cos(\theta(\sigma)$ are shown in (b). The extent of the `plateau'
  is reduced for larger $\delta$.  Similar curves are shown in (c) and
  (d) for ${\cal W}=10$, and $\delta = 0$, $0.1$ and $0.5$. (e), the
  integral $\Delta I$ (equation \ref{inte}) for $\delta =0$, as a
  function of $|{\cal W}|$. (f) The attenuation of the integral
  $\Delta I$ with respect to its value for $\delta = 0$ as $\delta$
  increases, for fixed ${\cal W}=10$. The linear part of the curve,
  for small $\delta$ can be fitted with a line of negative slope
  $Q_\delta$. (g) The slope $|Q_\delta|$ as a function of $|{\cal
    W}|$. (h) The critical value $\delta_c$ for which the integral
  $\Delta I$ attenuates to $10\%$ its value at $\delta=0$, as a
  function of $|{\cal W}|$.}
\label{fig:quasi}
\end{figure}

In case the condition (\ref{eq:stati}) is not satisfied, we still have
to check for the existence of terms $(m,\nu,k)$ which, albeit
non-stationary, exhibit only a small variation of the phase
$\theta(\sigma)$ over the period of the homoclinic transition. Such
terms will be called \textit{quasi-stationary} and they can be
selected from the remainder by the following procedure: neglecting the
slowly varying factor $[s_a(\sigma)]^{m-1}$ in 
Eq.~(\ref{eq:melnikint}), and factoring out a constant phase
$e^{i(\theta_0+{\cal N}\pi)}$, important quasi-stationary terms are
those for which the integral
\begin{equation}
\Delta{\cal I} = \int_0^{2\pi} 
\cos\left({\cal N}(\sigma-\pi) + {\cal W}\ln\tan(\sigma/4)\right) 
\label{inte} 
\end{equation}
has absolute value above a small (arbitrarily chosen) threshold
$\mu_0$. Consider for a moment the approximation ${\cal W}\simeq
const.$. Since the inspected term is assumed not to be stationary (not
selected by the condition (\ref{eq:stati})), we have that ${\cal N}$
varies according to ${\cal N}\geq -{\cal W}/2$ for ${\cal W}>0$, or
${\cal N}\leq -{\cal W}/2$ for ${\cal W}<0$.  Different values of
${\cal N}$ generate different behaviors for $\theta(\sigma)$,
symmetric with respect to $\sigma = \pi$, as shown in
Fig.~\ref{fig:quasi}(a). Figure~\ref{fig:quasi}(b) shows the functions
$\cos\left({\cal N}(\sigma-\pi) + {\cal W}\ln\tan(\sigma/4)\right)$,
for the same frequencies $\sigma$ of panel (a). From the comparison of
the two plots, we see that the nearly flat domains of the curve
$\theta(\sigma)$ near $\sigma=\pi$, along with the sigmoid variations
at the two ends (in panel (a)) imply the formation of a plateau of the
curves in (b) accompanied by fast lopsided oscillations, which nearly
cancel each other in the integral (\ref{inte}).  The flatter the
function $\theta(\sigma)$ in the vicinity of $\sigma = \pi$, the wider
is the plateau of $\cos(\theta(\sigma))$. Since the dominant
contribution in $\Delta{\cal I}$ comes from the central plateau
of $\cos(\theta(\sigma))$, the maximum absolute value of $\Delta{\cal
  I}$ occurs when the slope $\theta'(\sigma)$ becomes zero at
$\sigma=\pi$. Hence, from Eq.~\eqref{thetaprim}, the maximum occurs
when ${\cal N}=-{\cal W}/2$. The length of the plateau is given by
$\Delta\sigma_p = 2\sigma_p$, where
$\theta(\pi\pm\sigma_p)=\pi/2$. From Eq.~\eqref{thsigpend}, we find
$\sigma_p\simeq (24\pi/{\cal W})^{1/3}$, and hence $\Delta{\cal
  I}_{{\cal N}=-{\cal W}/2} \propto {\cal W}^{-1/3}$, an estimate
verified numerically (Fig.~\ref{fig:quasi}(e)).

On the other hand, if ${\cal N}$ is `detuned' from the maximum value
$-{\cal W}/2$, the associated plateaus attenuate, leading to a
decrease of $\Delta{\cal I}$. Yet, some of these contributions can be
larger than minimum threshold considered for Eq.~\eqref{inte}.
Setting ${\cal N}=(\delta-1){\cal W}/2$, Fig.~\ref{fig:quasi}(f) shows
the attenuation as function of the detuning $\delta$ for fixed ${\cal
  W}$.  For small $\delta$, the attenuation is nearly a linear
function of $\delta$ with negative slope, $\Delta{\cal
  I}\propto\Delta\sigma_p\approx (24\pi)^{1/3}{\cal W}^{-1/3}
-(64/3\pi)^{1/3}{\cal W}^{1/3}\delta$. If we extend the straight line
with negative slope in Fig.~\ref{fig:quasi}(f) up to the point where
the line intersects the axis $\Delta I=0$ we find a critical detuning
$\delta_c\approx(3\pi/2\sqrt{2})^{2/3}{\cal W}^{-2/3}$ beyond which
the term can no longer be characterized as quasi-stationary. Actually,
$\delta_c$ computed as above underestimates the true value of the
detuning, since (i) the curve $\Delta{\cal I}$ has a tail extending
only asymptotically to zero (i.e. as small as it may be, the
contribution of a quasi-stationary terms is never exactly zero) and
(ii) the slope found by linear fitting of the left part of the curves
$\Delta I$ vs. $\delta$ for various values of ${\cal W}$ shows that
the power law estimate of the slope $\propto{\cal W}^p$ yields an
exponent substantially larger than $1/3$ for values of ${\cal W}$ well
below unity (Fig.~\ref{fig:quasi}(g)).  On the other hand, a numerical
evaluation of the dependence of the critical detuning $\delta_c$ as
function of ${\cal W}$ (Fig.~\ref{fig:quasi}(h)) yields a law
$\delta_c\propto {\cal W}^{-q}$, with $q\approx 0.8$, i.e., slightly
larger than the theoretical estimate $q=2/3$. Taking into account all
these considerations, we formulate a heuristic criterion for
quasi-stationarity, namely:
\begin{eqnarray}\label{eq:quasta}
   &~& \mathrm{The\,\,term\,\,defined\,\,by\,\,}(m,\nu,k)
    \mathrm{\,\,is\,\,quasi-stationary\,\,\,} \\
    &~&\iff {\cal N}\cdot{\cal W}<0\,\,\mathrm{and}\,\, 
    |{\cal N}|<(1-\delta_c){|{\cal W}|\over 2} \nonumber
\end{eqnarray}
with $\delta_c=\delta_{c0}|{\cal W}|^{-0.8}$, where, by numerical
fitting, $\delta_{c0}\simeq 3$ for an adopted attenuation factor
$0.1$, or $\delta_{c0}=4.2$ for an adopted attenuation factor $\sim
0.01$.

The conditions (\ref{eq:stati}) and (\ref{eq:quasta}) are derived by
considering the upper branch of the separatrix solution
$\theta(\sigma)$. For the lower branch we have, instead,
$\theta(\sigma) ={\cal N}\sigma- {\cal W}\ln\tan(\sigma/4)$, hence we
obtain the same conditions for stationary of quasi-stationary terms,
but with the inequality ${\cal N}\cdot{\cal W}>0$ instead of ${\cal
  N}\cdot{\cal W}<0$. Also, the above analysis, based solely on the
behavior of the phase $\theta(\sigma)$, allows to identify stationary
or quasi-stationary terms for $|{\cal W}|$ arbitrarily large. It is
important to recognize that the quantity $\Omega = k.\omega_*= {\cal
  W}(|A|\beta\epsilon)^{1/2}$, represents the divisor associated with
the remainder term $(m,\nu,k)$. Thus, we may further restrict the
selection of remainder terms by retaining only those passing the
stationary or quasi-stationary criterion, and simultaneously
satisfying an upper threshold for the divisor value, say $|\Omega|<1$.

\begin{figure}[h]
  \centering
\includegraphics[width=1.0\textwidth]{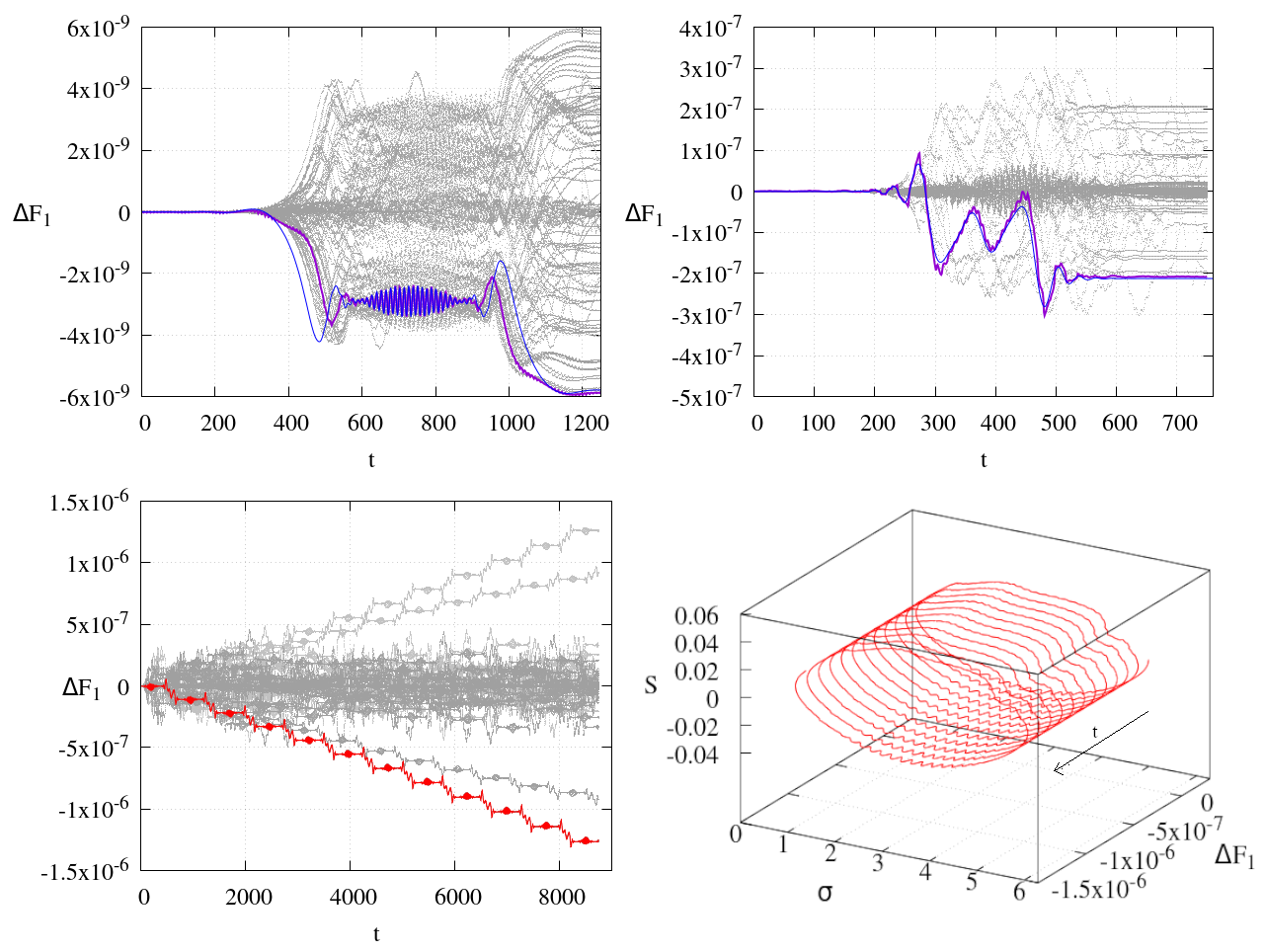}
\caption{\textit{Top:} Evolution of $F_1(t)$ for a swarm of 100
  trajectories with initial conditions very close to the hyperbolic
  torus at the simply resonant point $I_*$ same as in
  Fig.\ref{fig:arndif}, but for $\epsilon=0.003$ (left) or
  $\epsilon=0.01$ (right). The blue curves show the fitting to one
  trajectory of the swarm using the Melnikov integrals
  (\ref{eq:dftmel}) for only those remainder terms selected as
  stationary or quasi-stationary.  \textit{Bottom:} A ballistic orbit,
  drifting continuously in the same direction along the resonance, as
  depicted for the evolution $F_1(t)$ (left) or $F_1\times$ the
  pendulum variables $S,\sigma$ (right).  }
\label{fig:drift2}
\end{figure}
Figure~\ref{fig:drift2} shows the main result obtained by selecting
only the few terms ($\sim 1000$) of the remainder passing the criteria of
stationarity or quasi-stationarity.  Swarms of 100 trajectories with
initial conditions very close to the hyperbolic torus at the simply
resonant point $I_*$ same as in Fig.~\ref{fig:arndif}, but for
$\epsilon=0.003$ (top left) or $\epsilon=0.01$ (top right) for a very
small time ($T=1200$ and $T=700$ respectively), corresponding to the
time required for the orbits to complete the first homoclinic
transition along the pendulum, according to the approximative formula:
\begin{equation}\label{eq:talpha}
T_\alpha={1\over\sqrt{A\epsilon\beta}}\ln(32A\epsilon\beta/||R^{opt}||)~
\end{equation}
This formula is the same as Eq.~(\ref{eq:pendper}) used in section 2,
setting the pendulum energy as $\varepsilon=A\epsilon\beta$, with the
coefficients $A$ and $\beta$ obtained from the simply-resonant normal
form (section 3). As discussed before, showing numerically computed
original values of the adiabatic action $J_F(t)$ for these
trajectories provides no information, due to the deformation
effect. Showing, however, the same variable at optimal order by use of
the transformation (\ref{lietra}) makes clear the jumps along
homoclinic transitions exhibited by these trajectories. In particular,
we distinguish how the random distribution of the initial phases
results in a stochastic spreading of the actions
$F_1(t)=J_F^{(r_{opt})}(t)$ (with $r_{opt}=10$ in the left panel, and
$r_{opt}=7$ in the right panel), in a way qualitatively similar to the
one observed at Fig.(\ref{fig:ranphase}) in Arnold's model. The bottom
left panel extends the calculation in the case $\epsilon=0.01$ up to a
time $t=9000$. At this time the trajectories have undergone $~13$
transitions. The jump $\Delta F_1$ in every transition shows the
behavior of a random walk with size $~10^{-7}$. Thus, most
trajectories spread over an interval $(-(13)^{1/2} 10^{-7},(13)^{1/2}
10^{-7})$. However, we distinguish also rare trajectories which move
in `ballistic' motion, i.e., drifting systematically in the same
direction.  These are the fastest moving trajectories, with speed
bounded by an estimate which is the closest possible to the absolute
bound provided by the Nekhoroshev theorem. 
\footnote{A variational method to compute such fastest drifting trajectories 
in a priori unstable systems is proposed in \cite{zhang-11}.}
Note, finally, the excellent representation of the jumps by the semi-analytical 
(Melnikov) approximation (blue curves) using only the remainder terms selected 
by the stationarity or quasi-stationarity criteria.

For more rigorous statements on the (quasi-)stationary phase
approximation method see~\cite{GEP}.

\section*{Acknowledgments}

The authors acknowledge the project MIUR-PRIN 20178CJA2B "New frontiers
of Celestial Mechanics: theory and applications". C.E. acknowledges the support 
to this project by the H2020 MSCA ETN Stardust-R (GA 813644).

\end{document}